\documentclass[5p, times]{elsarticle}
\usepackage{multirow}
\usepackage[utf8]{inputenc}
\usepackage[english]{babel}
\usepackage[dvipsnames]{xcolor}
\definecolor{intnull2}{RGB}{214,214,214} 

\journal{Computers \& Security}

\begin{document}

\begin{frontmatter}

\title{A Survey on Analyzing Encrypted Network Traffic of Mobile Devices}
\tnotetext[mytitlenote]{This  work  was  supported  by  Center for Artificial Intelligence and Robotics Lab. DRDO India.}

\author{Ashutosh Bhatia${}^a$, Ankit Agrawal${}^a$  Ayush Bahuguna${}^a$, Kamlesh Tiwari${}^a$, K. Haribabu${}^a$, Deepak Vishwakarma${}^b$}
\address{${}^a$ Dept. of  Computer Science and Information Systems, 
Birla Insitute of Technology and Science Pilani, Pilani Campus, Jhunjhunu 333031, Rajasthan, INDIA\\${}^b$ Center for Artificial Intelligence and Robotics, Defence Research and Development Organisation (DRDO), Bangalore, INDIA}  

\begin{abstract}
    Over the years, use of smartphones has come to dominate several areas, improving our lives, offering us convenience, and reshaping our daily work circumstances. Beyond traditional use for communication, they are used for many peripheral tasks such as gaming, browsing, and shopping. A significant amount of traffic over the Internet belongs to the applications running over mobile devices. Applications encrypt their communication to ensure the privacy and security of the user's data. However, it has been found that the amount and nature of incoming and outgoing traffic to a mobile device could reveal a significant amount of information that can be used to identify the activities performed and to profile the user. To that end, researchers are trying to develop techniques to classify encrypted mobile traffic at different levels of granularity, with the objectives of performing mobile user profiling, network performance optimization, $etc.$ This paper proposes a framework to categorize the research works on analyzing encrypted network traffic related to mobile devices. After that, we provide an extensive review of state of the art based on the proposed framework.
\end{abstract}


\end{frontmatter}


\section{Introduction}
   The immense growth in the area of communication technology has been noticed in the last decade. In a short span of time, mobile devices become an integral part of our daily life. Nearly 45.04\%  of the world population (3.50 billion)  use smartphones \cite{web1}.  It is predicted by Statista that by the year 2023, this number will reach near to the present world population. The ways of accessing the Internet all over the world have been changing; desktop computers/laptops are no longer the preferred way to access the Internet, mobile devices own a significant amount out of total internet traffic. Currently, around 50\% of total internet traffic is being generated only through mobile devices, and it is expected to grow more as new technologies, \emph{e.g.}, 5G, will become more and more of a reality. As forecasted by the Cisco Visual Networking Index, 77 exabytes traffic per month is expected to be generated worldwide by mobile devices by 2022, and such amount of traffic would be around 20 percent of global IP traffic \cite{31Cisco}. The Ericsson Mobility Report states that this number nearly doubled between Q1 2018 and Q1 2019, and predicts mobile traffic to reach 131 exabytes per month by 2025 \cite{31Ericsson}.
 
    There are various user-friendly behavioural mobile applications such as Facebook and WhatsApp that are frequently used and generates a significant amount of internet traffic. According to the Statista 2019 report \cite{web2}, 2 billion users are accessing WhatsApp every month, and 1.3 billion users are accessing Facebook messenger.  A large fraction of Internet traffic originating from smartphone devices carry users' personal and behavioural information. A primary concern regarding these platforms is data privacy since most of the content is personal. External agents can tap the information flowing through the Internet at various levels by employing suitable data capturing techniques. For example, an attacker can install a malware in the targeted device and remotely command it to fetch sensitive information, including network traffic, without the user's knowledge. Hence, Most of the mobile applications implement encryption mechanism while transmitting the data over the Internet to protect the privacy of the user. Additionally, peer-to-peer applications such as eMule and BitTorrent  have also recently added encryption capabilities in their transfer mechanisms to avoid detection of its presence.

    Moreover, authorities may want to analyze the mobile traffic in the interest of citizens' safety and national security. Furthermore, social media has become one of the popular platforms for ease communication. Even though social media helps in connecting with friends and family, the activities performed in social media have started affecting national security. So there is a need to have legal agencies/authorities who can keep an eye on the suspected activities performed by users over social media platforms. Such agencies are not alone in the desire to analyze mobile network traffic. Furthermore, Internet service providers also perform NTA to discover the usage of network resources, identify the users and application types, and use this information for various controlling and monitoring purposes, including billing, quality of service, security \emph{etc.} \cite{33} 

    In general, Network Traffic Analysis (henceforth  referred as NTA) refers to applying various methods for making inferences from the network traces (collected at different points and different layers in a communication system) about the type of traffic, users, and type of devices corresponding to the traces. In literature, the researchers have broadly taken two different approaches to perform NTA for mobile devices. The first one is using inferential methods that preexist for classical Internet traffic, and customizing them to work for Internet traffic pertains to mobile devices.  The second approach is to design new techniques specific to mobile Internet traffic, considering the properties that differentiate it from traditional Internet traffic.
    
    Naive web traffic classification schemes commonly use transport layer port numbers to identify the applications, because well-known applications run on a specific port number. However, such schemes are not suitable for classifying mobile traffic, as most mobile applications use the application layer i.e., HTTP/ HTTPS, for communication purposes. \cite{34}. Another approach is Deep Packet Inspection (DPI) \cite{35}, which is used to classify the network traffic based on the payload inspection of TCP and UDP packet data. In order to do that, DPI searches for the signatures made up of the characteristics features of the applications.

    NTA is a challenging task to perform over encrypted data as the payload is encrypted, and it is not easy to extract the information by just inspecting the payload. However, researchers have observed that some factual information can be revealed from the encrypted payload to identify the applications and its users associated with a particular traffic flow \cite{37}. Machine learning and deep learning-based traffic analysis techniques are found more suitable to extract the traffic features and classify the encrypted network traffic \cite{aceto2019mimetic}. Countermeasures against statistical approaches to traffic analysis are not straightforward. A common underlying assumption is that padding is an efficient countermeasure against statistical analysis \cite{26}. Although some success has been shown in protecting browser generated traffic \cite{27} and the efficacy of traffic morphing \cite{28}, none of these techniques may be effective against robust analysis \cite{29}. 
    
    Mobile application platforms protect users' privacy by encrypting the data, making analysis hard. It may be crucial, for many reasons, to analyze such encrypted traffic and make useful inferences about the nature of the traffic and the mobile user. The ultimate aim of an organization (enterprise, government agency, Internet service provider) to perform encrypted traffic analysis targeting the mobile device, could be to support, Intrusion detection system, develop intelligence against cyber terrorism, profiling of mobile users, forensic analysis for cybercrime, quality of service provisioning in the network and network behavioral management \cite{cao2014survey}. As a consequence, several researchers have recently started investigating the techniques of performing NTA explicitly for the traffic of mobile devices. 
    
    In this paper, we provide a comprehensive literature review to investigate state of the art work on analyzing the encrypted mobile traffic using machine learning techniques. The paper makes following explicit contributions. 

\begin{itemize}
\item We categorize each work according to three different criteria: \\
\noindent 
1) objective of the analysis; \\
2) methodology adopted for ground truth construction, \emph{i.e.}, data generation, capturing and tagging techniques and; \\
3) technique to perform the classification, i.e., machine learning algorithm, input to the ML model, and feature selection. 

\item We discuss in detail the salient differences between the traditional web traffic, \emph{i.e.}, traffic generated using desktops/laptops/servers and web traffic targeting mobile devices. Such differences restrict the use of existing analysis techniques available for traditional web traffic without any modification. 

\item While discussing each work with one of the identified objectives, we provide an insight into the dataset used and the methodology adopted for classification. We also discuss possible countermeasures to thwart mobile traffic analysis and provide meaningful insights about challenges and pitfalls related to the topics that have been investigated and identify possible future research directions. 
\end{itemize}

We believe that our work will help researchers to explore the research gaps in this field and stimulate new research trends.

The rest of the paper is organized as follows. In section \ref{difference}, we explain the difference between mobile and non-mobile Internet traffic characteristics to justify why many researchers are working on the NTA of mobile devices explicitly.  Section \ref{applications} discusses several potential applications of performing NTA by adversaries, various federal agencies, and network administrators. In Section \ref{background},  we discuss the  widely used security protocols, and the ways of accessing information from encrypted network traffic.  In sections \ref{tax}, we elaborate upon our classification framework to categorize the existing works. After that, in  section \ref{litsrv}, we discuss the current situation in the field of NTA targeting mobile devices based on  proposed classification framework. In section \ref{counter}, we discuss the potential countermeasures discussed in the literature to thwart the encrypted traffic analysis. Section \ref{conclusion} concludes the paper by discussing a few potential research challenges in this field.

\section{Mobile vs Non-Mobile  Traffic Characteristics } \label{difference}
In this work, we refer the Internet traffic in two categories, 1) Mobile traffic: the traffic of mobile devices such as smartphones and tabs, and 2) Non-mobile traffic: the traffic associated with the traditional way of accessing the Internet such as using desktops and laptops. In this section, we discuss how the Internet traffic characteristics vary between mobile and non-mobile traffic to understand the differences and performing analysis of the traffic belonging to these categories. 

As far as Internet traffic is concerned, we restrict ourselves to client-server based communication and keep peer-to-peer communication generated by torrent clients or TOR browsers outside the scope. Further, the traffic in both the categories is divided in two sub-categories: \textit{browser-based} and \textit{app-based} traffic. The browser-based traffic refers to when individuals use their smartphones, tablets, desktops, or laptops to view online content via thin clients like Internet browsers. Whereas, the traffic generated by the thick clients, \emph{i.e.}, individual applications running on the device, is referred to as app-based traffic. 

The mobile-traffic and non-mobile traffic differ in many ways, and hence, the NTA techniques and related challenges. In the following, we highlight salient features of Internet traffic that differ between mobile traffic and non-mobile traffic and thus make mobile traffic analytic separate from non-mobile traffic analytic. 

\subsection{Protocols (Transport and Application)}

While categorizing the network traffic based on the transport layer and application layer protocols, it has been found that the significant part of the traffic generated by both device types is either TCP or UDP based. However, the amount of TCP traffic is much more than UDP. In the view of comparing traffic, the amount of UDP traffic generated by mobile devices is much less than the non-mobile devices \cite{gember2011comparative}. 

In consideration of application protocol, most of the traffic generated by both belong to web protocol, i.e., HTTP and HTTPS. It has been found that HTTP protocol is the most widely used protocol in the mobile app-based traffic by examining the HTTP header using user-agent strings for each device. Most mobile applications take help from some standard libraries for communication purposes, and these libraries add some fields in the user-agent-strings, which helps to identify certain mobile applications \cite{maier2010first}. Applications in the Apple devices use some standard libraries for communication purposes like the Apple CFNetwork library, which adds its version number and name to the end of user-agent strings. User-agent strings make it easy to create some patterns/signatures to identify and classify the HTTP protocol used in the network. Mobile devices generate more HTTP traffic than non-mobile devices \cite{maier2010first}\cite{gember2011comparative}. It has been observed that 82\% of HTTP traffic is generated by non-browser applications in mobile devices, compared to 10\% of non-mobile \cite{gember2011comparative}. HTTP-based video streaming mobile traffic is accounted for 42\% of total mobile HTTP traffic, whereas it is around 23\% for non-mobile devices. 

Another parameter used to compare mobile and non-mobile traffic is the distribution of HTTP object sizes. On average, HTTP object size downloaded by mobile devices is larger than other devices \cite{maier2010first}. Browser is the most popular application in mobile phones. Another observation is that the application size downloaded by non-mobile devices is larger than mobile phones.

HTTP protocol usage can be easily identified with the help of patterns created with the help of the user-agent string. However, It is not easy to detect non-HTTP traffic as other application protocols like POP does not add any required information in the user-agent string. Another characteristic of network devices is their IP TTL, whose default value is different in popular mobile devices. For example, the default TTL of the window is 128, and the TTL value for Macs is 64 \cite{maier2010first}.

\subsection{Traffic Flow Characteristics}

Another way to compare mobile and non-mobile traffic is to use the characteristics of the traffic flow generated by them. In general, traffic flow can be defined as a sequence of packets sent from source to destination during a specific interval. For example, the sequence of packets that are exchanged over a TCP connection can be considered a flow. For non-TCP traffic, such as UDP,  a flow can be defined as a sequence of packets having the same value of specific packet header fields: Source IP address, Destination IP address, Source port, Destination port, and the name of the protocol.

We use the number of packets per flow, flow size, flow duration, and the flow rate as flow characteristics to discuss the difference between mobile and non-mobile traffic. 

\begin{enumerate}
\item Number of packets per flow: The number of packets in a traffic flow generated by mobiles is much higher than non-mobile traffic flow. By considering several different types of devices and applications, the authors in their study \cite{lee2011study} found that 70\% of the flow in non-mobile traffic contains less than ten packets, whereas  60\% of the flow in mobile traffic contains less than ten packets.

\item Flow Size: Flow size can be defined as the summation of each packet's size in a flow. It has been observed that the flow size is smaller in the traffic generated by mobile devices than the traffic generated by non-mobile devices \cite{lee2011study}\cite{gember2011comparative}. Similarly, the ratio of download and upload traffic generated by non-mobile users is 2.9:1, whereas this ratio for mobile users is 5.9:1 \cite{afanasyev2010usage}.

\item Flow Duration: Flow duration can be defined as the duration between the first packet and the last packet in a flow or the duration between the connection make-up and break-up time. According to studies, the flow duration is shorter in the mobile-traffic than the traffic generated by non-mobile devices \cite{afanasyev2010usage}\cite{lee2011study}\cite{gember2011comparative}. In less than 50\% of the mobile generated internet traffic flow, the flow duration is less than 1 second, whereas, in less than 45\% of the non-mobile traffic, it is less than 2 seconds \cite{lee2011study}. A large portion of Internet traffic is attributed to network-based applications. With the consideration of a few specific applications, it has been found that the flow duration of mobile traffic is five times shorter than non-mobile traffic. IMAP and POP protocol based email traffic has shorter flow duration in non-mobile devices traffic, and SMTP based email traffic has shorter flow duration in mobile devices traffic \cite{gember2011comparative}. 

\item Flow Rate: Flow rate can be computed as the flow size divided by the flow duration. Both mobile and non-mobile devices have the same median flow rate of 10Kbps, but only 10\% of mobile flows are slower than 1 Kbps compared to 30\% of non-mobile flows \cite{gember2011comparative}. The transfer rates of the different classes of devices vary during the time of trace periods. The median rates for non-mobile users are 512 B/s and 128B/s for mobile users. On average, the transfer rate of non-mobile users is higher than mobile users \cite{afanasyev2010usage}.

\end{enumerate}

\subsection{Background Traffic} 
In addition to interactive traffic generated by users who perform particular UI activity in a mobile application, mobile applications also generate background traffic for various purposes, such as contacting their corresponding server to receive updates, maintaining their current state, or synchronizing with cloud services. For example, a newsreader app may generate a lot of background traffic as it periodically polls a server for the latest news. This situation is not very common in mobile web traffic where the background traffic generated by the web browsers is minimal. A study conducted by Time Stober et al. [16] reveals that 70\% of mobile traffic is background traffic, whereas only 30\% of mobile traffic is interactive. The study also revealed that patterns in this background traffic highly dependent upon the type of applications installed on the mobiles and also on device configurations. Hence it needs to be considered distinct from background networking processes on desktops. This variation in traffic patterns across multiple mobile devices, due to the presence of background traffic, can be effectively utilized to fingerprint these mobile devices. On the other hand, different mobile applications sometimes have similar background traffic as they may be built upon similar software libraries or application program interfaces (APIs that generate similar traffic irrespective of the mobile application in which they have been used. This kind of similarity in background traffic further complicates the analysis of the traffic.

\subsection{Services and Application Identification}
With the combination of HTTP hostname, one of the fields in HTTP message format is content-type, which allows identifying the type of services accessed by the clients. After grouping the information getting from the content-type field of HTTP header and analysis of the initial part of the HTTP body, we can classify the mobile applications, multimedia content i.e., audio and video, and the images.  Octet stream content subtype is used in 86\% of mobile application type data and 51\% of non-mobile application type data \cite{gember2011comparative}. RSS feeds, and MPEG-4 coding is another common application subtype for mobiles, and shockwave and adobe flash are common application subtype for non-mobiles. There are 185 different application subtypes available accessed by non-mobiles, whereas 58 application subtypes are available for mobiles. This massive difference in the availability of application subtype shows that non-mobile devices run a great diversity of applications. The traffic analysis for classifying the application is performed by seeing the first packet of each flow based on the protocols and port numbers. In order to do that, they build a mapping between the MAC address and IP address assigned to a device \cite{afanasyev2010usage}.

\subsection{Network Usage}
With network usage, we intend how much traffic applications impose on the network. A client application generates traffic during active periods. The criteria to consider a client to be active for a reporting interval is when the client sends at least one packet per second during the interval. One interesting observation is that mobile users are active only for a short period i.e., 40 to 80 seconds in an hour. Network usage also depends on the weekdays or weekend. It has been observed that mobile and non-mobile users' network usage is varying during a whole day on an hourly basis \cite{afanasyev2010usage}. A considerable variation is found in the activities done by mobile users than non-mobile users. Non-mobile users performed more activities than mobile users during the late afternoon.

\subsection{Different Websites}
A mobile phone's prominent property is its mobility, as it is smaller in size than a desktop computer. To keep the screen size in mind, many companies have an option to redirect their customer to a mobile-specific website or use another approach called responsive design which adjust the website layout according to the screen size. While accessing the internet, some mobile-specific dimensions, such as device name, device type, carrier network, and mobile browser, are inserted in traffic in the form of user-agent strings. In contrast, desktop computers are only concerned with operating systems and web browsers. 

\subsection{Identity Association:}
Associating identity with non-mobile-traffic is hard as the Internet could have been accessed from anywhere, like personal or office computers or cyber cafes. For mobiles, unique users can be identified via appropriate attributes like IMEI number or device model number in mobile devices, which are more resilient identifiers than cookies, which can be deleted. Further, due to mobile carrier contracts, people are often locked into using the same device for multiple years.
 
\subsection{Browser-based vs App-based mobile traffic} 

Now we discuss salient differences between browser-based mobile and app-based mobile traffic that makes network traffic classification and analysis of mobile applications different from browser-based mobile traffic. Traffic generated through Internet browsers mostly use standard application layer protocols such as HTTP and security protocols such as TLS, whereas the traffic generated by mobile applications depends on the nature of applications and their implementation. Another critical property that makes the nature of browser-based mobile traffic different from app-based mobile is the length of sessions and the level of activities inside the sessions. In general, the browser-based mobile traffic is likely to have a more extended session than app-based mobile traffic as mobile users frequently close the applications when they are not using it. Session teardowns and timeouts also impact analysis; session lengths are longer for websites and short, ranging in seconds, for mobile apps, which usually focus on a streamlined, service-oriented experience. 
 
\section{Applications of Network Traffic Analysis (NTA)} \label{applications}

The critical question is, why would someone be interested to know that a particular user or group of users used which application or performed which activities on their mobile phone. Internet traffic can be tapped anywhere in between the path from its source to the destination and to perform NTA. Tapping the traffic could be either a Law Enforcement Agency (LEA) intending to perform NTA to deal with issues such as investigating a possible cybercrime, identifying the source of cyber terrorism, building threat intelligence for the country, and performing cyber forensics to investigate a criminal or non-criminal case. Furthermore, social media helps connecting with friends and family, but the activities performed in social media have started affecting national security. So there is a need to have LEAs who can keep an eye on the suspected activities performed by users over social media platforms. Such agencies are not alone in the desire to analyze network traffic, Internet service providers also perform NTA to find out the usage of network resources and to identify the users and application types, and use this information for various controlling and monitoring purposes, including billing in ISPs \cite{rezaei}, quality of service, security, \emph{etc} \cite{33}. In addition to LEA and ISPs, the network administrators of an enterprise may also be interested in performing NTA on incoming and outgoing traffic for a variety of network management tasks. In the following section, we discuss the example applications of NTA.  

\subsection{Applications for Attackers (Cyber Criminals)}
Traffic analysis is considered as a severe threat to the network users and network itself as network traffic is open to access for all \cite{kausar2019traffic}. Attackers can make use of the information about a network/device/application in an endless manner. Here, we discuss a few ways in which an attacker can misuse the private information of the victim. 

The personalized phishing attacks may be launched by the attacker if he is aware of the activity recently performed by the victim over the Internet, such as visiting a job search portal or using a health application. If a smartphone is being used for home automation purposes, knowing the actions performed over the mobile phone can help the attacker to get the information about the presence of the owner of the mobile. The attacker performs an attack using traffic analysis techniques in order to infer the webpage visited by the user on the users' mobile phone \cite{kausar2019traffic}. Attackers can misuse the traffic features extracted by traffic monitoring. Attackers can find vulnerabilities in a network by just analyzing the network traffic and using it to his/her advantage. Attackers can identify the user's identity on a social media website like Facebook through traffic analysis attacks \cite{trujillo2019traffic}. 

Attackers can also use NTA techniques to observe the behavior of a social media user and then misuse that information by making a fake profile. Adversaries monitor the traffic generated from instant messaging applications like WhatsApp, which leaks sensitive information about their users \cite{bahramali2020practical}.

NTA can help the adversary in identifying the operating system of a mobile device. This OS identification can be considered as a starting point for an adversary in order to perform further attacks on mobile devices \cite{ruffing2016smartphone}. No mobile operating system provides 100\% security. As a result, each one has a list of vulnerabilities present in Common Vulnerabilities and Exposures (CVE). If an adversary identifies an OS with its version, then the adversary can exploit the vulnerabilities present in the operating system to perform more effective and advanced attacks.

Ad providers can be considered as a threat for mobile devices because ad providers perform analysis on the traffic generated by the ads in order to get the user’s Personally Identifiable Information (PII). Ad libraries are being used to leak many types of PII, i.e., gender, age, \emph{etc.} However, ad providers are not able to make a full user profile from the leaked information. Adversary correlates the information by exploiting the UDIDs in the ad traffic generated by many ad providers and trying to build a full user profile \cite{stevens2012investigating}. To prevent the leakage of PII, blocking the number of packets carrying PII or substituting the personal information with bogus data can be considered as a countermeasure against NTA.

\subsection{Applications for Authorities (LEAs)}
Social networking websites like Facebook, Instagram, Twitter, \emph{etc.} have provided ease of electronic communication and have become one of the platforms for sharing information. However, these platforms are also being used to post illegal content, rumors, spreading fake news, harassment to others by making fake profiles, and other illegal tasks. The common crimes committed using social networking websites include cyberbullying, stalking, hacking, frauds, fake profiles \emph{etc}. NTA can help government agencies to detect and prevent such threats by identifying the activities performed by people and who are doing such crimes. NTA can be used to perform network behavior analysis (NBA) to monitor unusual activities such as spreading rumors, posting illegal information \emph{etc.} performed by users over social media \cite{guan2014design}. Thus NTA helps in analyzing the post and reply information of a user. NTA can also help in finding the Internet usage pattern of a mobile user. The authorities can use the information retrieved from the usage patterns to profile a user for his/her gender, age, profession, \emph{etc.} This information would further help them to look for any unusual and suspicious behavior \cite{naik2020know}. Using more profiles of users, one can try to establish an association between them by carefully looking at their time-line of the online activities performed by them over the mobile device. For example, if we find a similar trend for traffic flow for two persons, we can suspect them to be talking to each other. Also, by analyzing their other activities, we can predict their social conduct, which can help in tracking unsocial people.

\subsection{Applications for Network Administrators/ISPs}
NTA helps network administrators (NAs) or ISPs in the process of network planning and management to improve the quality of network services, network behavior management, and the security of the network. In general, we divide the applications of performing NTA by the network administrators or ISPs into two dominant categories Network Management and  Network Security.

\subsubsection{Network Management}
One of the critical goals of NTA is in monitoring the application performance \cite{37}, utilizing, planning, and managing network resources and allowing them to make application-related policies within an enterprise network. NTA is used to perform network behavior analysis (NBA), which is one of the tools for network monitoring, and used to find abnormal behavior by analyzing the activities performed by users \cite{guan2014design}. NBA helps network administrators to acquire users' requirements, to optimize internet marketing and network management, \emph{etc.} There exist thousands of network applications that run over a network and consume different amounts of network resources. NTA allows the network administrator to find out the applications which are eating more resources than defined in their SLA (service level agreement). Another usage of performing NTA is to find out the popular websites visited by the users helping network administrators learn about users' behavior and other related information like visited website category, user's preferences, \emph{etc.} \cite{guan2014design}.

Additionally, the NTA may also help discover the unwanted traffic generated by employees inside an organization such as playing online games, watching movies, and other prohibited tasks during the working hours. This information can then be used to avoid wastage of bandwidth and other resources in the network. NTA also helps the network administrators in troubleshooting the problems efficiently. For example, it allows NAs to know the underlying causes of being network slowdown. One of the reasons is that most of the employees update their anti-spam software at the same time due to the high utilization of the link. The network administrator can make better strategies to get rid of such problems.

QoS provisioning in the network uses a priori knowledge about network characteristics to get the optimal performance from the network. Traffic reports also help the NA to predict the future load on the network. This information can be used by NA for efficient budgeting of the resources to provide better Quality of Services to the users.  Such analysis can be of use for various ISPs who can identify primary sources for bulk bandwidth consumption. For example, a provider whose services are being swamped by peer-to-peer file sharing might want to detect and selectively throttle problematic connections. This is a reasonably realistic example since P2P communications can be incredibly demanding on a network, and because protocols for file sharing between peers like BitTorrent are wildly used to download data in bulk. In general, knowledge of traffic statistics in advance can empower an ISP to accordingly control and redirect resources to optimize services in a utilitarian manner.  

\subsubsection{Network Security}
The Internet provides several services to users and ensuring the availability of services is a vital concern. Denial of Service (DoS) and Distributed denial of service (DDoS) attacks are found as severe threats against service accessibility. NTA is found one of the prominent solutions against such attacks \cite{37}\cite{niuapplications}. Attackers may inject trojans (a malicious code) in legitimate software to control the target machine without its user knowledge and send the user's personal and valuable information remotely to the attackers. Trojans are considered as a threat to privacy and data security \cite{chen2014method}. NTA helps detect the trojans. NTA gets the answers to hard questions like who did it, how it happened when it happened, and what are the things lost.

NTA helps NA to assess the behavior of network applications and even the whole network in order to detect or prevent the leakage of personal information. Thus it helps in preserving the privacy of users. NTA is used to perform network behavior analysis (NBA), which is found as one of the prominent solutions to identify security issues \cite{guan2014design}. NBA is used to prevent security threats by discovering the abnormal user behavior, thus allowing us to detect anomalies in the network. Hence, Traffic analysis allows NA to save time and money involved in the recovery process of the attack, which is about to occur.
Moreover, It helps in tracking the attackers and detecting the violations in the security policies.  If this situation occurs in the network, then NAs can block the packets having such information or replace that information with the bogus information. Another usage of performing traffic analysis is to detect the malicious behavior of an application trying to download or install the malicious code in the victim device.

\section{Encrypted Network Traffic Analysis} \label{background}
Due to the inherent properties of encryption in the network traffic, statistical and frequency spectrum analysis of packet data can be utilized to identify signatures and extract information related to the captured traffic of a targeted user. 
Before discussing works on classifying and analyzing the encrypted network traffic explicitly generated by mobile applications, it is essential to understand the generic methods of performing the analysis of encrypted network traffic, irrespective of the platform i.e., web traffic or mobile traffic. After that, we discuss the differences between encrypted web traffic in general and encrypted mobile traffic, which deter experts from using similar methods for both problems.

 The information flow through the Internet should be protected against various types of attacks to protect users' privacy. That is why most Internet applications use encryption techniques to secure the data while transmitting over the Internet. As a result, it becomes challenging to perform NTA over encrypted traffic. The Internet is designed as the layered architecture that allows us to implement different security protocols over different network layers. The most widely used security protocols include TLS implemented at the transport layer, IPSec implemented at the network layer, and other application-specific security protocols, e.g., SSH protocol for remote login application, message stream encryption (MSE) and protocol encryption (PE) for BitTorrent \emph{etc.} Here, we briefly discuss the two most widely used security protocols: Transport Layer Security (TLS) and Internet protocol security (IPSec).

 TLS is a security protocol that is implemented on top of TCP protocol at the transport layer and used to provide end-to-end security. TLS provides security to the application layer protocols \emph{i.e.} HTTP, FTP \emph{etc.}, which use TCP. TLS consists of four protocols: handshake, alert, change-cipher, and record protocols. However, major functionality is implemented in two protocols only, handshake and record. TLS begins with the initial phase called handshake protocol, which is responsible for the cipher suite (collection of cryptographic algorithms) negotiation, authentication of both parties, and finally, the key establishment by exchanging certain security parameters. Some unencrypted information is exchanged during this phase until key establishment. These keys with the negotiated cryptographic algorithms are further used in the record protocol to establish a secure session between the parties for transmitting the application data. TLS encrypts the payload but does not encrypt the transport layer protocol’s header. Therefore, it does not provide traffic flow confidentiality as a passive attacker can always see the TCP ports over which the communication is happening.

 IPSec is a security protocol that is implemented at the network layer to secure the communication between host-to-host, network-to-network, gateway, and host. IPSec is mainly used in the following: virtual private network, application-level security, routing security. IPSec consists of mainly three protocols: Internet Key Exchange (IKE), Authentication Header (AH) and Encapsulating Security Payload (ESP). IKE protocol is responsible for authenticating both parties, key establishment, and security association. The key establishment process includes the negotiation of cryptographic algorithms and required keying material. The data is securely transferred with the help of the other two protocols in the light of transport and tunnel mode. The transport mode does not encrypt the IP header of the original packet. Thus the transport mode of operations provides end-to-end security through a single tunnel. While tunnel mode encrypts the original packets' IP header, this mode of operation provides many tunnels between end devices.
 This feature of IPSec provides limited traffic confidentiality in the sense that an attacker can only get the IP address of tunnel end-points instead of actual source and destination IP addresses.

Most of the security protocols are implemented mainly in two phases: The first one is the initial phase where both parties authenticate each other and exchange the security parameters, other required information without encrypting them. The second one is the use of cryptographic primitives in order to secure the transmission. 

After performing traffic analysis, the information extracted from the initial phase of security protocols helps in distinguishing browsers, applications, and operating systems. The information extracted from the initial phase of security protocols i.e., the TLS handshake phase, helps estimate the user-agent of HTTPS client \cite{husak2016https}. User-agent is a field of HTTP request header whose value is used for client identification and identifies the operating system and application. Server Name Identification (SNI) is an extension of the TLS protocol. SNI allows a client to indicate which hostname it is trying to connect with, during the TLS handshake process. TLS protocol uses digital certificates i.e., X.509 certificates, in order to authenticate the parties. One party request a certificate from the other party and the other party sends it. One of the fields in the X.509 certificate format is a unique subject ID that uniquely identifies the peer to whom the certificate is issued. X.509 certificate helps one party in determining the other party involved in the communication \cite{holz2011ssl}. Traffic analysis helps in detecting man-in-the-middle attacks with the help of SNI and certificate both. By performing traffic analysis, the information extracted from the X.509 certificates determines both parties identifiers, which helps in uniquely identify the server involved in the communication and even helps in detecting server changes and malicious software \cite{37}. Hence even though it is difficult to extract the information from encrypted traffic, it provides some valuable information.

Another objective of NTA could be recognizing the standard cryptographic protocol used in the application. Accounting for the number of packets exchanged during the initialization phase, distributions of packet sizes, and their structure the underlying protocol can be identified easily. PACE \cite{38}, Cisco Network-Based Application Recognition (NBAR) \cite{39}, nDPI \cite{40} are some of the open-source classification libraries, written in C and primarily based on pattern matching and statistical analysis, that can recognize standard (TLS, SSH or IPSec) as well as non-standard security protocols (Bittorrent and Skype) from captured network traffic (encrypted or unencrypted). Having information about the application layer protocols in use can be constructive in further narrowing specifics of the application related to the captured network traffic. However, to get granular information about the captured traffic e.g., identification of the application protocol or activities on a web site, analysis of the encrypted transport phase is required.

Most of the internet traffic is encrypted as network applications use encryption schemes to secure the transmitted data over the Internet. Two approaches can be found to extract the information from the encrypted data. The first approach is to decrypt the encrypted traffic and get the information, but for decryption, the secret key is required, not publicly available. The second approach is to extract information from encrypted traffic only. Various NTA techniques can be applied over unencrypted data packets in order to extract the information. One such technique is Deep Packet Inspection (DPI), which is used to classify the network traffic by looking at the headers and payload of the packets. However, DPI is not suitable for performing packet inspection over encrypted data \cite{sherry2015blindbox}. 

Encryption hides the content, but it does not change the semantics of traffic flow characteristics \cite{krishnamurthy2013privacy}. As a result, quite an ample amount of information is left for performing NTA. Only encryption schemes cannot stop let obtaining information by performing NTA. So such NTA methods are needed, rely on the statistical feature of the traffic instead of the actual content. Statistical-based techniques can perform and classify the characteristics of encrypted traffic and are less affected by encryption than the DPI  \cite{37}.

\begin{figure}[t]
\centering
\includegraphics[scale=.1]{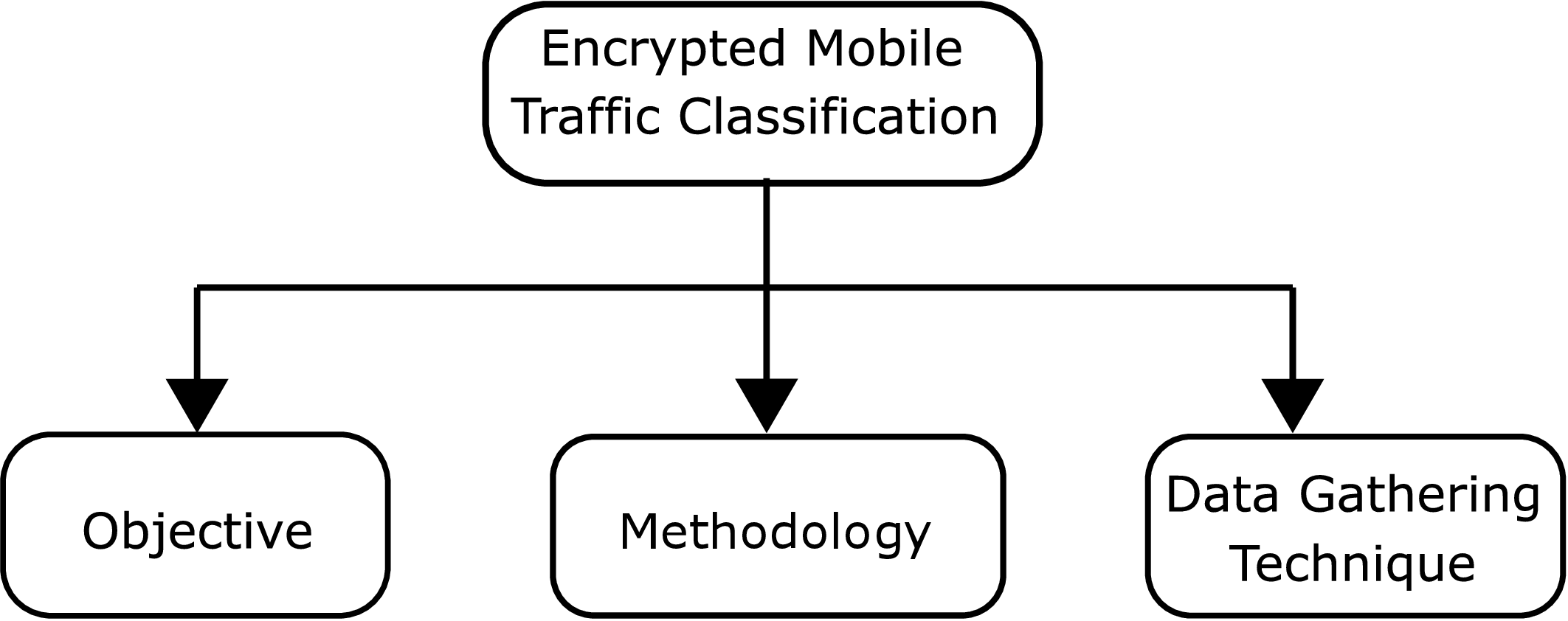}
\caption{Dimensions to classify the research on encrypted mobile traffic analysis}
\label{encryption}
\end{figure}

\begin{figure}[t]
\centering
\includegraphics[scale=.3]{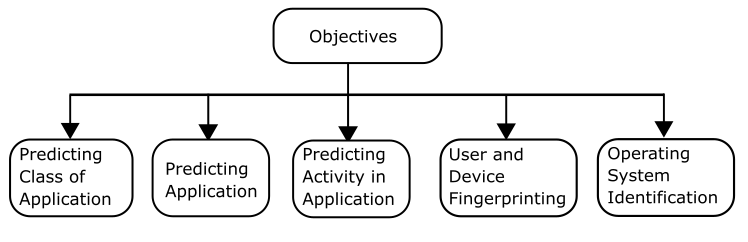}
\caption{Objectives for performing encrypted mobile traffic analysis}
\label{objectives}
\end{figure}

The methods of encrypted network traffic classification can be broadly classified into two categories payload-based and feature-based. In payload-based methods, the packet structure is crucial for identifying the application protocol, whereas feature-based methods leverage protocol flow characteristics for identification purposes. For payload-based classification, the same open-source tools used for security protocol identification can be used to identify the application protocol, as many applications in their initial phase follow a strict structure, just like the security protocols. Therefore, the same pattern matching approach can be used.

Machine learning and deep learning-based traffic analysis techniques are found more suitable to extract the traffic features and classify the encrypted network traffic \cite{aceto2019mimetic}. Over the last decade, a large amount of research has been conducted to design feature-based classifiers using supervised or semi-supervised machine learning techniques as a second category of methods, especially targeting the security protocols such as TLS, IPSec, SSH, Bit-Torrent and other similar protocols. This survey considers only those papers that use learning techniques to perform NTA with specific objectives. These techniques mainly focus on the features extracted from the traffic flow, i.e., flow characteristics, traffic behavior, and patterns.

In the following, we provide two examples to demonstrate how performing NTA using statistical techniques can reveal several critical privacy-related information about the user even when the traffic is encrypted.

 Website fingerprinting is one of the applications of traffic analysis techniques. NTA attacks to the encrypted HTTP traffic can reveal the website identity accurately. Encryption schemes provide secure tunnels to hide the browser activities from eavesdroppers, making it difficult to understand the data. However, they do not hide the traffic flow characteristics like packet length, the number of packets, timing, and packet directions. Such information leads to traffic analysis attacks, which can reveal the identity of websites \cite{herrmann2009website} \cite{cai2014systematic}. Attackers may find useful information regarding websites (URL, the content of the Website) implementing Privacy-enhancing technologies (PET) by performing traffic analysis. The attacks on Website fingerprinting exploit the HTML pages' structures and sizes, including the objects of HTML pages. Attackers observe the following properties of an encrypted connection: packet size and direction, and inter-arrival time to perform website fingerprinting attacks \cite{herrmann2009website}, revealing password in SSH logins  and many more.

SSH is considered a secure remote login protocol that provides confidentiality and authentication. However, it consists of two weaknesses, the way it uses the padding mechanism and the sending of a separate IP packet after each keystroke. The first weakness reveals the original packet size, and the second weakness leaks the information about the inter-arrival timing of packets. The first weakness can be exploited by using statistical techniques to find sensitive information such as the password length, and the timing information can be used to reveal the characters what the user typed \cite{song2001timing}.  These two weaknesses can lead to other severe attacks. The server sends one or more IP packets in reply against receiving each packet sent by the client after each keystroke. In the traffic signature attack, the attacker can easily find when the user is typing a password if there is no echo packet.

\section{A Taxonomy for Encrypted Mobile Traffic Classification and Analysis} \label{tax}
\begin{figure*}[t]
\centering
\includegraphics[scale=.6]{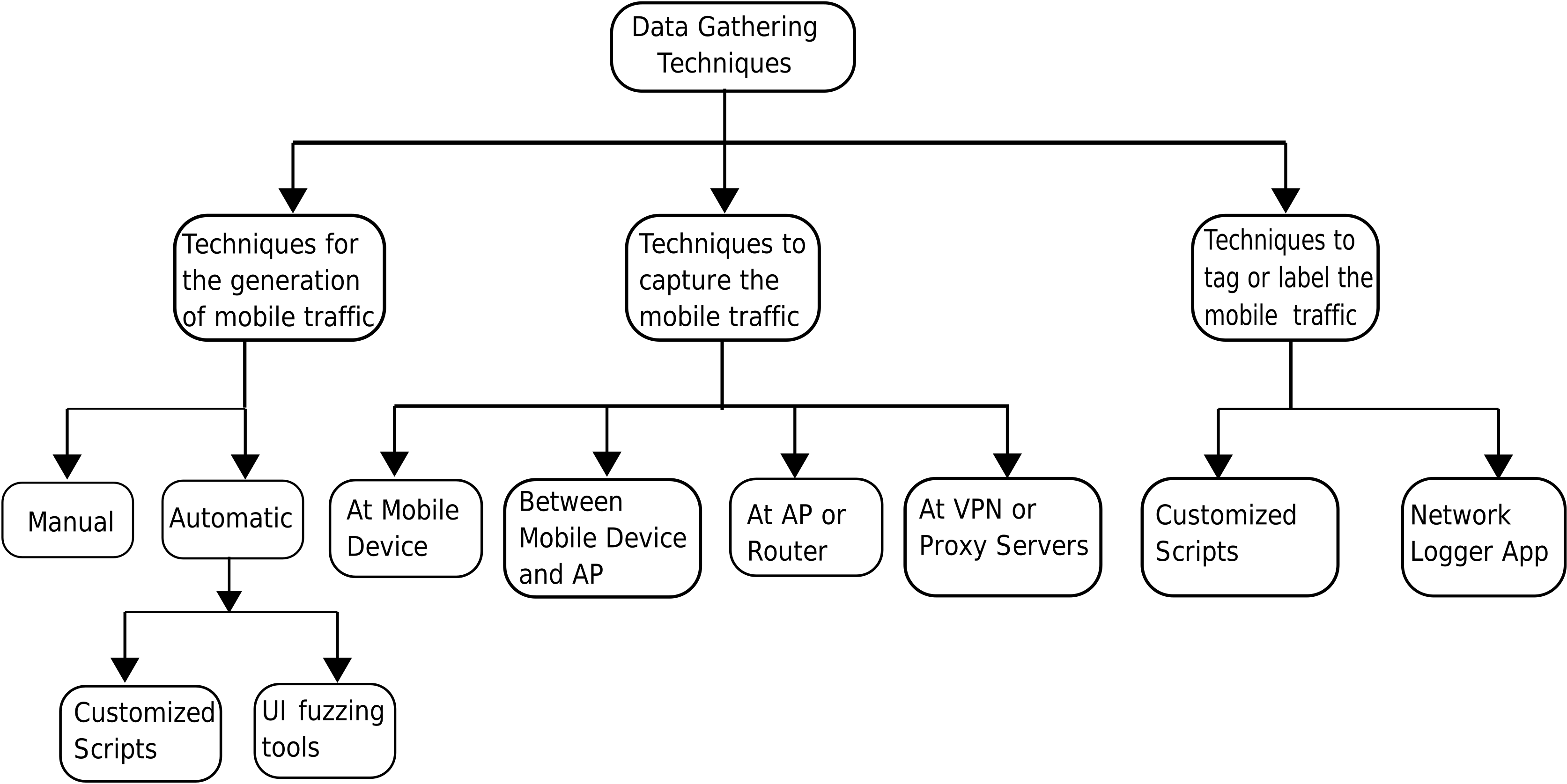}
\caption{A classification of data gathering techniques for encrypted mobile traffic }
\label{taxfig}
\end{figure*}

As described in Fig. \ref{taxfig},  we propose a  classification framework as the taxonomy to categorize existing works on NTA of encrypted mobile traffic to have a clear understanding of state-of-the-art in this area of research. The classification is done on various levels. At the topmost level of the hierarchy, existing research works can be distinguished based on their objective of performing traffic classification and analysis, techniques employed to gather the data, and finally, the approach or methodology applied to achieve these goals.  Moreover, we have further classified the data gathering techniques into a multilevel hierarchy, to be able to see a subtle differentiation between the related works. Finally, we provide three different dimensions that can be used to classify the approach or the methodology taken by existing works on performing NTA of encrypted mobile traffic.

\subsection{Objectives of performing analysis}
Based on the objective of performing the classification and analysis of encrypted mobile traffic, a particular mechanism can be classified into the following four categories. 
\begin{enumerate}
\item Predicting application class: Current mobile applications can be broadly divided into categories such as instant messaging, music, videos, gaming, mail, and social media applications. Besides the encrypted traffic generated by these applications, statistical artifacts of the communication (size and inter-arrival time distribution) turn out to be characteristic and do not change after performing encryption. This allows traffic analysis through developing classification models that can identify underlying implementations by matching temporal features. However, mechanisms to identify the class of application of encrypted traffic are generic in the sense that they can be used for both mobile and non-mobile traffic. 

\item Detecting a specific application:  Only knowing the category or class of the application is not sufficient to get the relevant details about the mobile users. Additionally, an adversary might be interested in knowing the mobile application responsible for the captured traffic. Inferring application usage of mobile users can reveal much personal information about their interests, hobbies, online shopping traits, and the name of service provided by obtaining financial, entertainment, and medical services \cite{53}. For this, the analysis has to be performed with higher granularity to predict the application used by the mobile user. A significant portion of research on analyzing the encrypted mobile traffic falls in this category. 

\item  Detecting specific activity in an application:
The applications installed on a smartphone can potentially reveal the traits of its user. However, such information may not be sufficient to know the specific actions performed by the user on her mobile. 
A user may perform different activities in a mobile application. For example, when using a social networking application like Facebook, the user can perform different actions such as sending a message, browsing profiles, and interacting through posts. Inferring the specific activity performed by a mobile user on their mobile device reveals more information about the user than only knowing the application. As a result, some efforts in literature pursue encrypted traffic analysis with an additional focus on predicting the specific activity or action performed by a mobile user. Much of this work is based on the assumption that the time series data produced by a particular activity of an application are usually different from those of other activities, where these time series record flow features like packet sizes and the transmission or reception time of packets belonging to that activity.  

\item User and device fingerprinting:  An critical requirement with the profiling of mobile users is establishing a mapping between the mobile device and its user's identity. The works with classification and analysis objectives, as discussed before,  do not explicitly address this issue. They can only classify traffic but cannot associate it with the identity of the person to whom this mobile traffic belongs. The task of establishing a certain identity with the mobile device to which a captured traffic belongs is called \textit{mobile fingerprinting}. The fingerprinting is done at various levels, such as identifying the underlying OS,  MAC address \emph{etc.}  of a user device.
Device fingerprinting (aka canvas fingerprinting, browser fingerprinting, and machine fingerprinting) is a process used to identify a device (or browser) based on its specific and unique configuration.

\item OS Identification: OS identification is a part of device fingerprinting. However, It is important to consider OS identification separately as OS identification in several contexts found valuable. For example, Specific OS is restricted to use in the enterprise network for security reasons. OS identification may help in following the same restriction. OS fingerprinting is also used to detect tethering.
Moreover, adversaries find OS identification, a starting point to perform further advanced attacks. OS fingerprinting is an integral part of penetration testing, which is used to find the vulnerabilities in the devices and networks and design and implement better security controls mechanisms. OS fingerprinting helps NAs in monitoring the number of devices and classifying the device roles. There are following certain traffic features being used in the literature in order to detect the OS: TTL and Identification field of IP header, window scale size and timestamp option in TCP, and others such as boot time and clock frequency. OS classification is performed by extracting the information from the sniffed packets' header.
\end{enumerate}

\subsection{Data Gathering and Tagging Techniques}

A large fraction of work on analyzing encrypted mobile network traffic focuses on identifying the type of traffic among multiple categories, considering it to be a classification problem to be solved using statistical or machine learning methods. The ground truth construction is a crucial task to develop a learning model based on any machine learning algorithm. In the context of classifying encrypted traffic with various objectives, it involves multiple tasks such as capturing of network traffic traces,  cleaning of captured traffic to remove the noises,  separation of entities (possibly a particular traffic flow)  to be classified from the cleaned traffic traces and labeling of these entities for training and testing purpose.  In the following section, we discuss various data gathering techniques and tagging that have been collectively applied by various research works. After that, we discuss a few frameworks proposed by some researchers to generate, capture, and tag the mobile traffic.

The primary requirement, while proposing a method to analyze encrypted mobile traffic, is to have sufficient data to train and test the proposed models. One approach is to use the datasets that are already available with cellular network operators.  Originally, these are collected by network operators for billing and monitoring purposes. Due to the sensitive information that they hold, network operators have been very cautious about sharing them with other parties. By applying anonymization and information aggregation schemes and allowing compliance with privacy-preserving regulations, operators have shared such datasets for research and development purposes in the past. However, after applying anonymity, such datasets are only useful to extract the trait information about a community instead of a particular user or to perform user and mobile fingerprinting. 

An alternative solution is to generate data ourselves and use it to develop classification models. A majority of works have taken this approach. However, the techniques employed to generate, capture, and tag data vary across these works. It is noteworthy that some techniques for generating mobile traffic may be more suitable for particular objectives of performing analysis than others. 

\subsubsection{Generation of Mobile Traffic}
One way to perform user simulation i.e., to generate the mobile traffic so that it can be captured and further analyzed, is to take help of volunteers allowing their network traffic to be captured. The advantage of this approach is the possibility of capturing the mobile traffic that got generated due to  interaction  of real user with the applications installed on the  mobile. However, finding such volunteers in bulk is difficult as well time-consuming; it could take months to collect data sufficient to build reliable machine learning based classification models. Secondly, it could be possible that these volunteers may, consciously or not, defer from the usual realistic actions they perform on devices by either trying to suppress or exaggerate biases and favouritism between applications or action. Another challenging issue with this approach is the tagging of the network traces collected in this manner. Some sort of network logger application has to be installed in each volunteers device to identify the application responsible for a particular network flow present in the captured network traces. 

Another technique to generate the mobile network traffic is to  write programs capable of running the applications installed on a mobile automatically by connecting the mobile to a workstation or laptop. For instance, the Android development platform provides a command line tool called Android Debug Bridge (ADB) that allows a machine to manipulate a mobile device to taking various actions such as installing and opening of an application on the target  mobile. This approach is advantageous in many respects against the manual or volunteer-aided traffic generation approach as experiments can be performed in more controlled environments. Faster data collection, broader coverage of mobile applications and ease of labeling data are some of the benefits of using automated generation of traffic against manual methods. 

A challenge associated with automated generation approach of mobile network traffic is scripting application-specific actions on the device that can be used to uniquely identify the application. The network traffic generated by scripts for a particular application should ideally cover the entire spectrum of network flows that can possibly be generated by the application. This requirement is very much similar to the scenario when a newly developed application has to be tested. The developers  of the applications try various sequences of user actions (test cases) either on an emulator or on an actual device to test the application thoroughly before releasing consumer-ready versions. This technique of using a script to elicit responses from mobile applications by simulating user actions within the mobile application is known as UI fuzzing. These UI events are generated randomly to generate multiple network flows of the mobile application. One commonly used tool to perform UI fuzzing available as part of the standard Android SDK UI automation toolkit is \texttt{monkeyrunner} \cite{47}, which is able to capture traffic generated by applications once deployed. Frameworks such as Dynodroid \cite{48}, that use advanced UI fuzzing techniques, can be also used to aim for better results in terms of covering the number of flows of an application.

It is important to discuss why and when the models, built on a dataset which are generated in such an automated manner, would work to classify the unlabelled realistic traffic flow - the flows that have been extracted from mobile network traffic generated by the applications when a human user have used the services provided by the application in a legitimate manner. UI fuzzing tools are useful as classifiers traditionally do not use explicit behavioural characteristics such as click rates that would manifest only in human generated application traffic.

\subsubsection{Capturing of Mobile Traffic}

Various techniques to capture mobile network traffic can be categorized based on the exact location and device on the network where the eavesdropping of packets takes place.  

\noindent \textbf{At Mobile Device}
A packet sniffing mobile application such as tPacketCapture \cite{49} are used to capture the mobile generated traffic. Mobile operating system provides certain services like Virtual Private Network (VPN) which are used by such traffic capturing mobile applications. These applications use some standard file formats such as pcap in order to store the captured traffic. Such files are stored in the secondary storage of mobile devices or can be uploaded onto the server. Network traffic analyst can either collect them directly from mobile devices or download from the server for further processing. One of the  advantage of using such packet sniffing applications is that they can perform  packet capturing without using root permissions for the mobile.  However, this technique fail to capture the packet traces from many applications as certain applications do not work with VPNs. Also, it is observed that such packet sniffing applications capture only those packets that are sent when the application is running in the foreground. For this issue other packet capturing software such as \texttt{tcpdump} \cite{50} can be used  which is capable of capturing packets at lower levels, and hence, is not application dependent. However, such low level packet analyzers require root access of the mobile device to be able to intercept mobile traffic. Another issue associated with this technique of having packet sniffers at the mobile device is, to capture packets on a large scale, the packet sniffing application has to be installed on the mobile devices of hundreds of volunteers, which seems in feasible. This becomes even more harder when the root permission is required on devices.

\noindent \textbf{Between Mobile Device and Access Point or Base Station}
Another way to capture the information being communicated between mobile device and base station or access point is by placing the packet capturing devices in the proximity and putting them in promiscuous mode. However, only a few research works on analyzing the encrypted mobile traffic have taken this approach as the frames being communicated between mobile device and the base station (in 3G or 4G technologies) or the WiFi access point are usually encrypted. Such captured frames only expose the MAC addresses of the source/destination by completely hiding TCP/IP header information. Although the actual classification performed by most efforts does not directly rely on TCP/IP header information, this information greatly helps in building the ground truth of the dataset to be classified. In particular, TCP/IP header information is required to separate out various flows from the captured network traffic.

\noindent \textbf{At Access Point or Router}
The most common approach to capture network communications, adopted by a majority of research efforts on mobile network traffic classification is to sniff the packet traces at the intermediate routers which are responsible to forward the incoming packets in the direction of their corresponding destination, or by  directly placing the packet sniffer on the same device in which WiFi access point  is located.  For example, if mobile device is connected to  WiFi access point which is then connected to a workstation to extend the Internet connectivity available at the workstation. In this case,  packets which are coming from or destined to the mobile device can be easily captured by placing a network sniffer (e.g Wireshark \cite{wireshark}) at the workstation. However, with this approach, 3G/4G mobile traffic can not be captured. 

\noindent \textbf{At VPN and Proxy Servers}
Data generated on mobile devices can be captured at a centralized Virtual Private Network (VPN) server using IP tunneling. It involves creating a VPN server and configuring the VPN connection settings on the mobile devices that routes all incoming and outgoing traffic of the device. VPN server uses a virtual adapter just like a dial-up connection to facilitate the communication between the source and the destination. A packet analyzer co-located with the VPN server can be configured to listen the virtual adapter to capture packets going through VPN server. One difficulty with this approach is that certain operating systems like Windows, provide limited support for capturing the packets from dial-up adapters. This method for data capture solves the problem of installing packet analyzers or networks loggers in every mobile device, and also, the data from both WiFi or 3G/4G interfaces can be captured. However, the mobile devices need to be configured to use the VPN service.
 
A similar way to intercept mobile  traffic over the Internet is by using proxy servers. One such open source proxy server is Tinyproxy \cite{51}. It is a HTTP based proxy server daemon for POSIX operating systems. It can be used to intercept network data at a centralized server, much like VPN servers. The advantage of using Tinyproxy is that intercepted packets are accessible through packet sniffing utilities like Wireshark.  
 
\subsubsection{Labelling Captured Mobile Traffic}
After generating and collecting the trace of network traffic, an important requirement is to labelling various traffic flows present in the captured trace file so that the captured data can be used to train and test the prediction models. The process is also known as ground truth generation. Here a flow is considered as the entity or data point which needs to be classified. The  definition of the flow depends on objective of classification. For instance, all the packets transmitted or received by a mobile device over a TCP connection can be considered as a flow to be tagged and later to be identified using the developed model. Another approach is to divide a longer TCP connection which usually contains multiple data \textit{bursts}, separated in time by a fixed threshold duration where no communication has happened over the TCP connection. One can assume that a single burst in a TCP connection carries sufficient features to identify the  application associated with the connection, and can also assume different labels and predictions for each burst. It is hence crucial to remove unwanted traffic such as packets related to re-transmission due to network error or even the packets belonging to other applications running in the background. The average mobile OS usually runs a number of background processes whose communications impose load on bandwidth from time to time. This unwanted traffic could make the process of fingerprinting traffic sources difficult. Filtering the packets related to re-transmission is relatively simple as the re-transmitted packets contain the same segment number of the original transmission. 

Here we discuss a number of techniques commonly used to identify the application responsible for each network flow coming from the target with the purpose of tagging the flows accurately.

\noindent \textbf{Network Logging on the Target Device}
A network logger or socket logger application in mobile device can generate data logs containing information about the application associated with each packet sent or received by the mobile device. By using these logged data along with customized demultiplexing scripts one can tag the flows with their associated applications. One such network logger application is Network Log which is an open source application that once started on target device can identify the application responsible for each network flow coming to the target device. An issue with these applications is that they require the target mobile  device to be rooted which may not be a feasible all the time. It is to be noted that these network loggers are slightly different from packet sniffing tools such as \texttt{tcpdump} which cannot associate the process ID of the application with a captured packet as they do not understand the target mobile kernel level information. On the other hand, these network loggers monitor the target mobile's kernel and associates each packet flow with the UID of its controlling application. The socket log along with the socket pair information (four tuple) also includes the user ID and the state of each active TCP/UDP session. Each session is mapped with an application according to the relationship between UID and application, since each application is assigned a unique UID when installed.

In case the objective of the analysis is to identify the application class of the captured mobile traffic, instead of an individual application, a static mapping between the application and the class can be used to tag the application class of the flows present in the captured network trace. 

\noindent \textbf{Customized Scripts}
Tagging of the flows in the above manner (by use of network loggers and demultiplexing scripts to filter the traffic) does not work well when the objective is to precisely  fingerprint the activity inside an application performed by the mobile user. A TCP flow may contain multiple  activities and an application may choose the same TCP connection to send some backround traffic (a traffic that does not belong to any user activity) over which an activity is being performed. For example, the Gmail application may send and receive emails over the same TCP connection or it can start the process of syncing with the server using the same TCP connection that has been established to send an email. No readily available application or framework is available to generate, capture and tag the traffic associated with the activities performed by the use in an application. Scripts can be extended to help the process of tagging the captured data, by recording the start and end time of an activity to a log file. They can record such information as they themselves trigger the execution of user activties of an application on the mobile. Together with logs containing start and end timestamps of each activity, demultiplexing scripts, TCP socket information and the Server Name Indication (SNI) field available in the TLS header, it is possible to tag the flows with their corresponding user activity with very high accuracy. SNI provides an optional clear-text indication of the target server's address in the opening phase of communication, and allows TLS servers with credentials for multiple identities to select and transmit a client-requested identity instead of blindly providing a single identity. As an example, google  uses  SNI-enabled HTTPS servers to host multiple secure website certificate identities on the same IP address and port. Without SNI, HTTPS servers can only effectively serve one certificate identity per IP address. In practice this results in wasted IPv4 addresses which have non-trivial costs. SNI support is commonly enabled in modern web browsers. 

These methods often benefit from clustering techniques making it possible to label data en masse; as these techniques inherently peer into close relationships between datapoints, they facilitate labeling for a significant bulk of samples based on a reliably small \textit{seed} dataset. Hierarchical clustering and similar algorithms make it easier to manually label data, hence these scripts can also be used in a \textit{semi-automated} manner with human supervision.

\subsection{Prevalent systems for generating data}

One can simply use present open source methods for the purpose of constructing labeled databases such as Mobligt \cite{32}, which incorporates a VPN server that maintains TUN interfaces with devices having a client application installed. This application routes all traffic for the device through the interface using network address translation (NAT). Both client and server services log socket data to correlate application label with each session captured by the VPN. For each timestamped traffic trace, a labeling model on the server matches socket data with tuples of application and socket identifiers, as well as timestamps. 

DELTA \cite{30} is a proposed framework for mining configuration and event data on Android devices for research and monitoring purposes. A user creates installable \i{experiments} on a desktop GUI in the form of application instances that can be managed using a core managerial app. Each instance has a log creation and viewing tool as well as extensions which specialise in communicating readings from different sources. In this manner, each experiment can be initiated with the needed permissions so that the user does not need to manage them individually on a single logging tool. A total of 44 status can be logged, including sensor readings, screen and system states, and also statistics for network traffic and application events. The framework also has a web service that can be used to store and download logs on demand.

SystemSens \cite{systemsens} proposes a similar system, where a client app subsumes logging services and data from \texttt{/proc/net/dev} provide network statistics. A server application is responsible for collecting data from devices and maintaining a database using SQLite. An uploading service sends JSON data to an external server with storage and visual aids for analysis. The system however has limited polling options, where a constant polling rate needs to be hardcoded. 

Another option is to build upon programmable logging frameworks. An example for Android is Phonelab \cite{phonelab}, which is a popular open-source solution with already implemented modifications for network statistics and scan results, and similar options exist for iOS.

\subsection{Methodology} \label{methods}

A simple way to identify the application of a network traffic is to make use of transport layer port numbers assigned by Internet Assigned Number Authority (IANA) to the applications. In this method, to identify the traffic, a classifier has to simply look only for the TCP handshake messages (SYN packets) exchanged between the sender and receiver during the connection establishment, and therefore, this method can provide a real-time performance.  However, the port based
classification fails when applications are either not registered with IANA i.e., port numbers are not well known or port numbers are decided dynamically. Both the situations are very common for mobile traffic as most of the mobile applications use some ephemeral port to communicate with their server or multiple applications use the same
port number.  Finally, the port based identification can only be used to classify the mobile applications, but to identify the exact activity performed by the user in that application.

A commonly used technique to classify the network traffic without using the port numbers is to look for patterns or signatures of an application present in the payload of the packet, also known as “Deep Packet Inspection (DPI)”.  For example, the application header of a packet can be examined to identify whether it is an HTTP message that is being exchanged between a web client and a web server by looking for a regular expression for HTTP signature. Most of the Internet Service Providers (ISPs) use this technique as their packet filtering mechanism to reroutes or blocks the packets depending upon the payload being carried by them. However, breaching the privacy of the user and the limitation to perform payload inspection due to encrypted data are a couple of major reasons, due to which DPI is not a suitable option to perform network classification and analysis over encrypted mobile network traffic.

Another majorly used technique to identify the network traffic which is also applicable for mobile traffic analysis is to build the classification models based on the statistical properties of the captured mobile traffic.  This involves, formalization of relationships between the statistical properties of the traffic in form of mathematical equations.  Some of the relevant statistical properties of the network traffic which can work as the feature to distinguish between the various types of traffic are: packet size, flow- duration, inter-arrival time of the mobile traffic are considered to develop the classification model by the formalization of relationships between variables in the form of mathematical equations. he major issues with the statistical approach for developing the classification model is its incapability to handle large datasets and the human involvement while building the model. This led the researchers to develop and use machine learning based classification models for mobile NTA. Although the core of machine learning is again statistics i.e., making sense of data by extracting its statistical features, the machine learning-based models can handle the classification problem with humongous dataset having much more complex boundaries and uncertainties.

In this study, we survey only the works which have used machine learning, deep learning, and other related techniques to classify the encrypted mobile traffic with the objectives discussed earlier.


In recent years, machine learning has challenged already established methods in several analytic domains in terms of performance and robustness. The central driving idea behind such learning approaches to analysis is to replace the trade-off between accuracy and efficiency with one between \textit{training} and \textit{testing} computational resources. This is done by isolating data- and resource-heavy computation to training, and generating minimal data structures that encode the requisite behavior desired during application. The efficacy of a learning algorithm is evaluated on it's ability to both satisfactorily perform such encoding during training and perform under testing.

Machine learning algorithms prove to be really beneficial in traffic classification but currently the market for mobile applications is expanding and evolving very fast. There are innumerable sets of apps each with their own unique traffic characteristics. Such speedy growth and development of mobile apps is making it hard for analysis of network signatures to study their unique features and perform accurate traffic classification. For this reason we discuss deep learning, another approach to the problem that undercuts the importance of isolating optimal feature sets.

Deep learning aims at designing classifiers capable of deriving higher-order features from inputs even for complex patterns which can’t be studied . Unlike classifiers that optimize costs inferred from handcrafted features, algorithms in this subset of machine learning converge on features inferred directly from input data which can offer better, dynamic traffic classification for any varying and complex mobile traffic pattern. In essence, deep learning methods employ subsequent non-linear filters to derive mappings input and output spaces, with the \textit{depth} of the model as a hyper-parameter, where increased depth allows the model to approximate complicated solutions but requires more training computation. How such approaches handle the trade-off between training and testing resources is often decisive for application.

The pre-processing module in such a domain imposes various restrictions on the model. Once the traffic is captured the first discriminating factor is how traffic is divided into various traffic units on the basis of port or protocol parameters. The type of input data being fed to the traffic classifiers can be, for example, the initial $n$ bytes of payload in binary, the initial $n$ bytes of raw data or the informative data fields of first $n$ packets. One aspect common to all deep learning approaches is that the input provided to model has to be of fixed dimensions. A workaround for this is truncation of longer instances or padding shorter instances with zeros, and the model must appropriately \textit{learn} to ignore such variations on underlying patterns.

These restrictions come with significant payoff as deep learning models achieve high praise for their ability to identify patterns in noisy and dense data while being adaptable with other existing implementations, making them ripe for computer vision and signal processing applications \cite{dnnhmm, incv3, wavenet}. Rezaei \& Liu \cite{rezaei} discuss various implementations under the lens of encrypted traffic classification and the restrictions faced in achieving scalability for data and implementation specifications. They also discuss the \textit{recurrent} neural network (RNN) architecture, where input data are iteratively fed to sequentially connected, identical neural layer blocks, predicting an output sequence at each cell step. These blocks, each representing a vector of non-linear features of both input data as well as prior predictions, functions as a recursive circuit, and hence these networks take advantage of an effective artificial "memory", similar in essence to how sequential circuits provide computational memory. RNNs also offer a means to use variable length sequential inputs with deep learning methods. Such reinterpretations of the standard feedforward, Restricted Boltzmann Machine \cite{rbm} model for non-linear learning will be vital to make such an approach relevant and versatile enough for traffic analysis.

Though literature on methodologies and results is thriving and multiplying with the field progressing daily, yet wide-spread adoption of such methods still faces hesitation. One key reason for this is the necessity of adequate amounts of filtered data required by these models, making several such approaches off-limits to average user without means or expertise to collect and filter data. Even corporations with well-established database management systems demur from readily applying machine learning in public domains due to security issues regarding \textit{adversarial learning} that have been gaining attention in recent research, wherein malicious attacks intended to impede learning systems can hamper accuracy which are difficult to identify and correct. Hence methodologies in this field require specific scrutiny before considering adoption.

\begin{figure}[t]
\centering
\includegraphics[scale=.48]{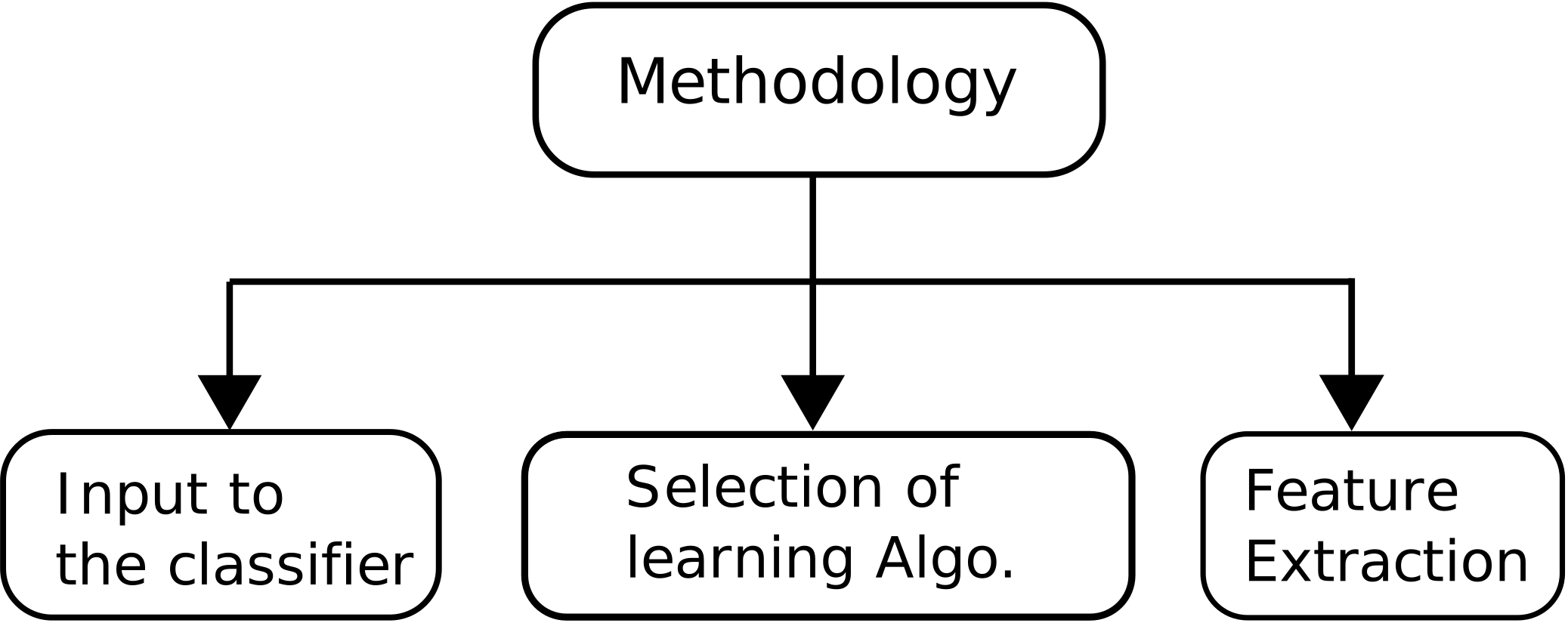}
\caption{Dimensions to classify the learning based methodologies that are used to perform NTA for mobile-traffic}
\label{methodologies}
\end{figure}

A learning approach can differ fundamentally on three facets, i.e., the data fed to the ML algorithm, how the features are selected from data and the classification algorithm (Fig. \ref{methodologies} In regards to our problem, these aspects are discussed briefly in the following points. 

\subsubsection{Input to the classifier}
All accumulated data for user traffic instances need to be pre-processed into formats compatible with the machine learning module. One way to allow such algorithms to make packet data legible for data analysis is to divide the network traffic sequences into certain discrete units called \textit{bursts} and \textit{flows}. The definition of bursts and flow varies from application to application. At a basic level a burst can be defined as sequence or group of packets only arriving at a certain time. A packet is said to belong to a different group or different burst when the arrival time of packet is after the specified burst threshold. So for a group of network packets in a burst the inter arrival time is shorter and within a specified burst threshold time $t$. Packets belonging to a single burst can have different destination address and source address, and hence need to be distinguished as separate flows - packets with same destination and port address in a burst are taken as one flow of network packets. Different applications starting TCP sessions simultaneously may send and receive packets belonging to the same bursts but will be considered under different flows. This definition of bursts and flow is not concrete and might not suffice all applications or user activities, for which we will need to define flows on a finer level. But we can conclude by saying that, in general, our classification framework will process data in the structured form of flows, dependent on network traffic. 
The only exception to this is in the case of side-channel data which, as we shall discuss in section \ref{litsrv}, can significantly bolster the performance of the algorithm but inevitably limit it's applicability due to the selective nature of data required and challenges in applying such methods at scale. These methods are especially powerful when applied in controlled environments where authoritative agents can assume or ensure the availability of build and system variables, but face difficulties in regards of responsible disclosure of information by users. Spreitzer et al. \cite{sidechan} describe paradigms for classification of algorithms exploiting side-channel information. An in-depth survey is provided on the basis of passive or active attacks that respectively expose either logical or physical variables and operate either remotely or from the vicinity of the target.

\subsubsection{Selection of learning Algorithm}
Now that we have defined the presentation of data to the algorithm in question, it is important to be clear about two broad categories of machine learning: unsupervised and supervised. The unsupervised class of algorithms works on unlabelled data aiming to identify hidden structure in input based on similarity measures like Euclidean distance either in the feature spaces of the problem and create clusters, or in conjured higher dimensional spaces. Such methods are useful when it is unclear how to perceive aggregated information, but it can be hard to evaluate any solution for the very same reason. On the other hand, supervised machine learning algorithms learn from previously labeled data to classify unseen instances. Labeled past instances can be used to increase accuracy of classifying unseen instances; for the objective of interest these class labels need to communicate the nature of the activity a person uses an application for during a sequence of bursts, to help the model distinguish between, say, email, IRC services, video streaming and so on.

Pertaining to how these supervised and unsupervised learning algorithms can be used for mobile encrypted traffic classification, an ensemble of methods can often be quite effective, for example unsupervised clustering and supervised classification. In this vein, clustering can be used at a preliminary stage to group and label similar flows in a burst using statistical parameters such as average flow separation time. Hence, clustering can provide a more legible format for training data that, when used together with appropriate supervised classifiers, can provide better results than na{\"i}vely training either on large blocks of network packets. Common algorithms used for clustering in various often follows agglomerative hierarchical clustering \cite{fahc, 41,10} or k-means clustering.

 The next module, the classifier, is trained with instances of mobile encrypted traffic flow labeled with application classes, from this learning a model can be built by generalizing our derived training dataset. Ideally, this dataset should have a large amount of instances with adequate variance between different datapoints for permissible fitting of the model on data. \textit{Cross validation} is a prevalent method to test learning algorithms for performance and repeatability, where the available data are divided into $n$ equal partitions and each partition in turn is used for testing a learning model trained on the remaining samples. This process is hence repeated $n$ times for each partition where usually $n \in [5,10]$ with a tradeoff between a sufficiently large sample of classifier evaluations and an adequately large testing dataset size, hence providing a meaningful average of performance metrics.

\subsubsection{Feature Extraction}
For the classifier to eventually to converge to accurate prediction of the classes, it is important to select a specific set of relevant features and  the rest. Using redundant features increased the costs involved in the storage of data and the computational resources required by the classifier. Here, the relevant field of study is dimensionality reduction. It helps in fast training of model, reduces complexity, improves accuracy and reduces over fitting. Existing feature selection methods are filter method which uses certain metrics to rate the features and select the best among those, although the type of ML algorithm used can itself limit the kind of features that are applicable. A few features commonly used are send and receive average inter-packet times, sent and received data size ratios \emph{etc.}. Moore et al. \cite{discr} elaborate on the subject with 249 possible traffic features that can be probed for different problem statements. A method predominantly used for feature selection is to rerun the classification experiments by only using selected subsets of features, converging on subsets which lead to maximum accuracy. It has been a common observation in all the works that removing features of little overall importance can lead to performance loss; features with misleadingly low variance that are often discarded during dimensionality reduction can be decisive for outlier points that are not adequately represented in training data. 


\section{Literature Survey} \label{litsrv}
In this section, we discuss existing works on performing NTA of mobile Internet traffic using learning-based techniques. The works on performing NTA of non-mobile Internet traffic, or the mobile traffic but with the objective other than objectives covered under the proposed classification framework, are out of the scope of this survey. We scrutinize the works under the study individually according to the proposed taxonomy. We group the existing works based on their objectives and discuss them together in this section.

\subsection{Application classes}
Al-Naymat et al. \cite{44} pursue discriminating between popular VoIP and non-VoIP applications like Skype, YouTube, and PayPal that were used to create a realistic dataset. Data were sniffed using Wireshark and then preprocessed into TCP flows for feature extraction classifier; packet length, cumulative bytes, and the times elapsed since the last packet and elapsed since the first packets of flows were four features deemed useful for classification. A single node was dedicated to intercepting and labeling traffic requests traffic for four nodes on a network monitored using Wireshark. The J48 implementations of meta.AdaBoost and random forests, as well as multilayer perceptrons classifiers, are used to classify the traffic into VoIP and non-VoIP classes. Boosting is used to improve the performance of J48, a binary decision tree classifier implementation of the C4.5 algorithm, while multilayer perceptron networks are a widely used class of feed-forward artificial neural networks (ANNs). Random forests prove to be more accurate and more resilient against noise. The results exhibit a true positive rate of 98\%.  The performance was quantified using measures such as accuracy, precision and recall, and confusion matrix. The area under the Receiver Operating Characteristic (ROC) curve is used to gauge the sensitivity and specificity methods. It has been observed in the paper \cite{al2016classification} that in order to classify the VoIP and non-VoIP traffic, meta.AdaBoost has shown the highest accuracy of 98.3007\% with the comparison of random forest and MLP classifiers having an accuracy of 96.6615\% and 84.166\%, respectively. All applied methods maintained high precision rates on all algorithms, save for multilayer perceptrons that lagged behind GTalk and Skype data performance.

The objective of Zhang et al. \cite{42} is to investigate the possibilities of information leakage related to user privacy by guessing the online activities performed by the users using inter-arrival time, packet size, and flow direction features extracted from data link layer frames. Traffic, which is considered to be of seven classes of activity from a single source application, is shown to vary considerably in different environments. Concurrent online processes are also considered in the experiment due to which traffic features of one application may be submerged by another application, further increasing algorithmic complexity. The model is limited to detect a maximum of two concurrent applications. A framework for hierarchical classification is developed that segregates traffic features using a decision tree of classifiers.  These features are calculated as packet-level statistical values of flows captured in different network environments. Two classifiers employing the radial basis function are used in the decision tree, RBF kernel SVMs and RBF networks, a reinterpretation of traditional ANNs where each neuron performs a radial distance metric around a point whose position vector converges during learning. The result shows that good accuracy is achieved that is robust to noise, where segregation into seven classes of activities is performed with around 80\% accuracy and 90\% accuracy when traffic is sniffed for five seconds and one minute, respectively.

Auld et al. \cite{bnn} uses a deep learning approach with a modification of standard ANNs called \textit{Bayesian neural networks}, in which Bayes theorem is iteratively used to calculate posteriors for a given prior assigned to weights for a network with hyperbolic tangent nonlinearities. The open-source tool \texttt{tcptrace} was used to generate statistics from TCP flow, and these data points were labeled semi-automatically. Traffic was reduced to flows, and content of each flow was used to match packet data with host knowledge, along with active labeling where the user generating traffic cooperates with the labeling process, making manual classification possible for millions of flows. The model achieves accuracy higher than 90\% for seven classes and an average accuracy of 95.3\% over the dataset, where system error rates were optimized by modulating a rejection rate based on entropic prediction confidence.

Lopez-Martin et al. \cite{rcnn} makes use of a recurrent neural network architecture, where they make use of the most popular archetype of RNNs called long short-term memory (LSTM) models. Feature vectors generated from incoming packets are sequentially presented as a matrix to a modular network, with multiple convolutional and LSTM layers where depths of these modules were varied. The best-performing model was evaluated on the 15 most popular classes in the dataset, and experimental accuracy was unanimously high; however, precision, recall, and hence F-measure were non-uniform between classes.

Bar-Yanai et al. \cite{knn} presents an approach to classifying traffic using a hybrid of \texttt{k}-means and \texttt{k}-nearest neighbours algorithms. A dataset was generated using the Endace tool \cite{endace} and labeled on a per-flow basis by a Cisco SCE 2020, a professional hardware \textit{service control engine} for network traffic classification and manipulation. During training, improved \texttt{k}-means clustering was used to isolate a list of cluster centers. A \texttt{k}-nearest neighbor algorithm was then applied on the set using this list of centers to generate clusters with no overlap, leveraging the low computational complexity of the \texttt{k}-means approach but greatly increasing the accuracy. The hybrid algorithm thus consistently recreates the accuracy of the \texttt{k}-nearest neighbors but boasts quicker learning as dataset dimensions increase.

Park et al. \cite{ga} apply a genetic algorithm (GA) to isolate optimal feature sets for boosting classifiers used to deduce application classes responsible for analyzed traffic. GAs are an evolution-inspired approach to learning that represent a population of possible solutions using a "genetic" encoding and competitively score and compare these solutions with each other and with "offspring" solutions generated in preferential iterative reproduction by mutation functions applied on the digital encoding, assuming a form of competitive natural selection. A reduced error-pruning decision tree classifier was best boosted by the approach, and outperformed boosted J48 and na\"{i}ve Bayes classifiers.

Lotfollahi et al. \cite{dpkt} present a Deep Packet scheme that leverages a one-dimensional convolutional neural network architecture for the ISCX VPN-nonVPN dataset \cite{iscx}, a data-link layer traffic repository labeled by application and class of service, mainly on VPN or non-VPN nature of the underlying service. The model achieved 0.93 F-measure and precision and 0.94 recall for application class identification.

\begin{table*}[h]
\centering
\caption{Identifying class of services}
\label{serviceTable}
\begin{tabular}{|c|p{1.5cm}|p{1.5cm}|c|p{1.5cm}|p{1.5cm}|p{1.8cm}|p{3cm}|}
\hline
\multirow{2}{*}{S.No} & \multirow{2}{*}{Ref.} & \multicolumn{3}{c|}{Data Gathering Technique} & \multicolumn{3}{c|}{Methodology} \\ \cline{3-8}
& & Generation & Capturing & Tagging & Input & \multicolumn{1}{c|}{Algorithm} & \multicolumn{1}{c|}{Features} \\ \hline
1. & { Al-Naymat et al. \cite{44}} & Manual & VPN & Manual &TCP Flow & Random forests & Packet length, cumulative bytes, latest and cumulative time intervals \\ \hline
2. & {Zhang et al. \cite{42}} & Manual & VPN & Manual &TCP Flow with IP filter& Multi-classifier decision tree & Bidirectional MAC-layer traffic features \\ \hline
3. & {Auld et al. \cite{bnn}} & Scripted & At access point & Semi-automated & TCP Flow & Bayesian neural network & Flow and packet metrics, effective bandwidth \\ \hline
4. & {Lopez-Martin et al. \cite{rcnn}} & Scripted & At access point & Automated & UDP \& TCP Flow & Recurrent convolutional neural network & Packet size, port information, timing and direction, timestamps, TCP window size \\ \hline
5. & {Bar-Yanai et al. \cite{knn}} & Automated (hardware tools) & At access point & Automated (hardware tools) & UDP \& TCP Flow & Hybrid (\texttt{k}-means and \texttt{k}-nearest neighbours) & Packet statistics \\ \hline
6. & {Park et al. \cite{ga}} & Manual & VPN & Manual & TCP Flow & Multi-classifier decision tree & Genetic algorithms \\ \hline
7. & {Wright et al. \cite{28}} & Manual & VPN & Manual & TCP Flow & Hidden Markov models & Inter-arrival time, consecutive packet sizes \\ \hline
8. & {Lotfollahi et al. \cite{dpkt}} & Manual & VPN & Manual & TCP \& UDP Flow & 1D CNN & Inter-arrival time, bandwidth \\ \hline
\end{tabular}
\end{table*}
\subsection{Application identification}
\begin{table*}[h]
\small
\centering
\caption{Application Identification}
\label{applicationTable}
\begin{tabular}{|p{.75cm}|p{2.2cm}|c|p{2.3cm}|c|p{1.9cm}|p{1.9cm}|p{1.9cm}|}
\hline
\multirow{2}{*}{S.No} & \multirow{2}{*}{Ref.} & \multicolumn{3}{c|}{Data Gathering Technique} & \multicolumn{3}{c|}{Methodology} \\ \cline{3-8}
{}&{} & Generation & Capturing & Tagging & Input & Algorithm & Features \\ \hline
1. & Yao et al. \cite{yao2015samples} & Scripted & On device & Manual & HTTP payload and application identifiers & Lexical context analysis (non-learning) & Application identifiers and payload \\ \hline
2. & Rao et al. \cite{7} & Scripted & SSL interception at VPN & Manual & Unencrypted HTTP flow & Host name matching (non-learning) & Host name, user agent string in http header \\ \hline
3. & Spreitzer et al. \cite{22} & Manual & At Mobile & Manual & Traffic traces & Jaccard similarity & None \\ \hline
4. & Qazi et al. \cite{6} & Scripted & Network loggers on SDN controller & Automated & TCP flow & Random forests & Packet statistics \\ \hline
5. & Taylor et al. \cite{8} & Scripted  & b/w Mobile \& AP  & Manual & TCP flow  & Random forests & Packet length statistics  \\ \hline
6.& Taylor et al. \cite{9}& Scripted & b/w Mobile \& AP & Manual& IP filtered TCP flow & SVM and random forests & Packet length statistics \\ \hline
7.& Aceto et al. \cite{52} & Manual & Logged by mobile services provider & Manual & Feature Vector& Ensemble classifier & Packet sizes \\ \hline
9.& Aceto et al. \cite{dltc} & b/w Mobile \& AP & Logged by mobile services provider & Manual & Transport layer biflows & Variations of CNNs, LSTMs & Packet size statistics \\ \hline
10.& Mongkolluksamee et al. \cite{11}& Manual & At VPN& Automated & TCP and UDP flow& Random forests& Packet \& Graphlet statistics\\ \hline
11.& Le et al. \cite{15}&Manual& At mobile device& Scripted & TCP flow &Linear SVM&Identifiers in HTTP flow\\ \hline
12.& Wang et al. \cite{1}& Scripted & b/w Mobile \& AP & Manual & Windowed traffic samples & Random forests & Packet statistics\\ \hline
13.& Watkins et al. \cite{19}&Manual& b/w Mobile \& AP & Manual& Time-stamped ICMP replies & Neural fuzzy classifier & Packet arrival delay\\ \hline
14.& Alan et al. \cite{24}& Scripted & At access point & Automated & TCP traffic & Multinomial na\"{i}ve Bayes classifier & Packet and burst lengths\\ \hline
15.& Shen et al. \cite{bigrams} & Scripted & At access point & Automated & TCP flow & Second order Markov chains & Application attribute pairs\\ \hline
16.& {Lotfollahi et al. \cite{42}} & Manual & VPN & Manual & TCP \& UDP Flow & 1D CNN & Inter-arrival time, bandwidth \\ \hline

\end{tabular}
\end{table*}

There are examples in the literature of non-learning models used for the problem, and we explore a few examples to compare with machine learning approaches detailed further.

Yao et al. \cite{yao2015samples} proposed a framework named SAMPLES to identify the mobile application for a particular flow by classifying the mobile generated traffic based on application ID and HTTP header information. It collects data from randomly selected applications from platforms like iOS and Android and builds a repository of application IDs and names for as many applications as possible, and this repository is maintained and queried for lexical matching. SAMPLES individually simulate applications that essentially deal in HTTP communication; flows are grouped as a flow set with each header having an identifier for parent app paired with an application identifier string. This string comprises three components, the identifier type, the payload, and any prefix/suffix of the string occurrence referred to as \textit{lexical context}. Rule sets are made by manual conjunction of the lexical context and fed to the application identifier engine, which is trained to identify the application for the flow. This framework was run on 15 million flows generated by 700k collective Android, iOS or Symbian applications, but is limited to only unencrypted HTTP traffic and is not a learning model.

Rao et al. \cite{7} present a platform named “Meddle” \cite{45}, which uses a software middlebox and virtual private network (VPN) in order to improve the transparency and to control the privacy leaks for the internet traffic generated from mobile devices. Meddle software employs VPN tunnels using the open-source StrongSwan project \cite{45} to isolate mobile traffic independent of device characteristics like operating system or carrier. Labeling is done manually by installing the app and then performing automated interactions with it for ten minutes before uninstalling it. This automation utilizes the Monkeyrunner app scripting tool \cite{47} to emulate user actions on the applications to derive realistic data points. When traffic arrives at the server, Meddle uses \texttt{tcpdump} to record traffic. SSL bumping or SSL interception \cite{46} techniques are used for decryption and accessing the payload of encrypted flows. In SSL interception, the client and the proxy (here, the VPN server) establish a connection while the proxy independently processes the certificate received from the target server after a request. Using the Online Certificate Status Protocol (OCSP), the VPN can verify the server certificate, regenerate a proxy certificate signed with the key for an installed CA certificate, and present it to the client. The proxy server decrypts outbound traffic, and can process it for use by applications on any layer. User apps are distinguished by matching hostname fields, and devices are identified using the user agent string contained in the HTTP headers. The certificates, server name indication, and DNS messages are also used to map flows, offering 92\% accuracy. There are both large as well as niche VPN networks for which this model can be applicable, but the approach does not make use of learning models to utilize better-labeled data flows since hostnames can offer services of several natures.

 Spreitzer et al. \cite{22} presents a client-side-attack able to guess the browsing behavior by exploiting the data-usage statistics for website fingerprinting even in the presence of TOR and defense mechanism i.e., SSH. The experiment was performed on multiple 3G devices with different browsers. \texttt{tcp\_snd} and \texttt{tcp\_rcv} as well as Android \texttt{/proc} files were used for usage statistics. An intersection-over-union score with the test trace is used to rank candidate websites based on signatures updated every 10 seconds with a sampling frequency of 50 kHz. The framework accurately classifies more than 95\% of traffic, both standard and that routed through Tor, for monitored pages.

 Qazi et al. \cite{6} is a learning approach in which network loggers are employed to fetch about 200 flows per application for 40 popular applications on devices on a software-defined network (SDN). The size of the first $n$ packets, the port number, and IP address ranges are singled out as features for classification. With the use of the OpenFlow (OF) protocol in the SDN, network traffic logging and detection of the corresponding application is automated, which simplifies constructing the test-bed. Traffic traces are communicated to an SDN controller that utilizes decision trees, the performance of which was then judged using F-measure. The authors achieved 94\% accuracy on an average for classifying the applications in SDN.

Taylor et al. \cite{8} approach uses side-channel data for identifying mobile applications. First, they aim at collecting network traffic data and generating a unique fingerprint of each app, which makes its traffic pattern different than the rest. This information is further used for the identification of apps of unknown records. The definition of traffic flow is based on the IP address of generated traffic. The data in this experiment were collected between the Wifi access point and the internet. Consumer actions were imitated by fuzz testing of the user interface to generate a dataset. After traffic is collected, flow separation is done based on the IP address. Traffic ambiguity detection is also performed to tackle traffic generated by third party libraries to help the classifier tackle noisy data. For each of the flows, three vectors i.e., size of incoming packets, outgoing packets and both, were considered. Features were generated using the statistical properties of each vector and then passed through the classifier. Ambiguous flows were neglected, and the remaining flows classified using random forests. For evaluation, the data collected are split into training and testing datasets. Results were reported over multiple datasets with performance on each being tested 50 times and then averaged. The accuracy was between 65.5\% to 73.7\%. The study suggests that operating systems and implementation details do not significantly affect fingerprinting, but the app versions have intense effects. The effect of noise on classifier performance was also studied, and ambiguity detection and classifier validation are claimed to make a sharp improvement in performance. 

Taylor et al. \cite{9} also showcase work towards accurately identify the smart-phone application in use, for which traffic is captured between the Wifi access point and the Internet using the \texttt{tshark} library, a terminal-centric implementation of Wireshark. Generation of tagged data required scripts used to simulate user action, which is separated into individual flows using the IP address. A total of 18 distinguishing features from the flows are used involving statistical properties such as series length, measures of central tendency, $n^{th}$ moments, and percentiles. For classifier design, six machine learning approaches are used, each approach using either an SVM or random forest classifier. These approaches considered using different classifiers for different apps or different flow lengths. The results outline a trade-off between classifying more flows at the cost of decreased accuracy and only processing flows, which promise higher accuracy of prediction at the cost of labeling a much smaller fraction of the data.

Aceto et al. \cite{52} employs an ensemble classifier using fusion techniques in conjunction with multiple base modules proposed for traffic classification. A dataset is initially formed from global mobile solution providers of actual users on Android and iOS applications. Network traffic is granulated into service burst, a set of packets within single bursts belonging to bidirectional flows on the same protocol. The features considered to be of interest were sizes of incoming packets and outgoing packets. Attention was focused is on isolating combining methods, which offered the most improvement over the results achieved instead by just using the best performing classifier. Classifier fusion techniques employed were Majority Voting, Weighted Majority Voting, Na\" {i}ve Bayes, Behavior-Knowledge Space method, WERnecke's method. Except for BKS, all the classifiers had separate benefits, with majority voting offering the best improvement on precision and F-measure while Na\" {i}ve Bayes classifiers showed marked improvements to recall and overall accuracy. Employing a restricted set of base classifiers is shown to increase the accuracy of the model. The work makes a strong case for applying ensemble classifiers that can offer different advantages depending on the scenario of practical use of the combined framework.

Aceto et al. \cite{dltc} also explore several deep learning approaches for the problem, testing variations of 1D and 2D convolutional neural networks, LSTM models, and hybrids, on three manually curated datasets of flows of transport layer communications of the device. The different classifiers are evaluated using 10-fold cross-validation and compared, where 16-layer 2D CNNs and LSTMs exhibit consistent performance. The study stresses generality. Hence the results focus on classifiers with unbiased inputs and show that deep learning approaches can achieve good accuracy without bias. Although they lose generality in the process, the methods can achieve a significant boost to accuracy with bias.

Mongkolluksamee et al. \cite{11} tackle the objective to identify the specific mobile application where traffic is encrypted, but mobile apps exhibit similar communication patterns and destination hosts. A dataset was constructed using 3G data traffic of apps running in the foreground of smart devices. Traffic for 30 minutes of use of five popular apps was captured in PCAP format and filtered to leave only TCP and UDP communication traces. To train the classifier, a combined vector of packet size and features were provided. Traditional packet features, which the study claims are not effective alone, are complemented with \textit{graphlet analysis}, which in network theory is used to study local structures and communication patterns in networks. To analyze statistical features, the Abacus algorithm \cite{54} is used, which divides traffic based on packet and burst size characteristics. The final feature vector provided to a random forest classifier comprised 35 shape-based features from graphlet information, and 24 packet features. When evaluating the model, the F-measure was found to be 0.95 when randomly sampling 50 packets in any 3-minute trace of the dataset. 10-fold cross-validation is used for testing sessions of each application's traffic trace. The work claims better identification of apps by combining network structure and communication pattern features.

Le et al. \cite{15} the objective is to monitor and analyze large traffic traces generated by a system named Ant-Monitor. The analysis aims to identify applications responsible for traffic. However, the focus is monitoring and analysis of coarse data collected by ISPs or intermediaries in the network and fine-grained data captured at the users' end to facilitate passive monitoring for consumers. The system consists of 3 components Ant-Client, Log-Server, and Ant-Server. Ant-Client is an Android client application installed in the user's device that app implements a VPN service. A virtual interface communicates all outbound traffic from any Android application, controlled by Ant-Client running as a background process. The app allows the user to toggle the VPN service and selectively monitor particular apps. A logging module in the client tunnels the packets or packet headers to Log-Server and maps the packets to the applications they belong to. Analysis of traffic is done at the user side completely without privacy concerns. The dataset collected in the study is from student volunteers at UC Irvine, who installed Ant Client on their phones for two months. Both Wifi and 4g traffic for popular applications were captured and labeled by the application name by the framework. Supervised classification on TCP traffic flows of about 70 applications was performed using 84 largely statistical network-level features on upstream and downstream flows. Tenfold cross-validations are used to train and test a linear SVM model, which exhibits a 70\% F-measure.

Wang et al. \cite{1} use side-channel attacks for sniffing encrypted traffic under the assumption that there is prior knowledge or access to the targeted device's MAC address. A signature is generated for each app using five minutes of captured traffic. Every collected traffic burst is analyzed in a windowed format for a given sliding window length parameter. Twenty packet-level statistics are nominated as features, half each for incoming and outgoing traffic, and used by a random forest classifier. The work stresses on feature selection as crucial for the task; experiments were repeated with different input vectors to find the optimal feature set. Sources of noise were also singled out with considerable monitoring time of the framework that introduced noise and concurrent applications running in the phone, limiting classifier accuracy.

Watkins et al. \cite{19} developed a mobile device's resource monitoring method which can remotely detect the different type of currently active mobile applications ( I/O intensive, CPU intensive and non-CPU intensive applications) by capturing ICMP replies, calculating inter-packet spacing and using Neural-Fuzzy Classifier (NFC). Analysis of network traffic for approximating computational power of mobile devices is performed to study the effects of resource throttling on network traffic. Network traffic is intercepted on a monitoring laptop between a wireless AP and targeted mobile devices. \texttt{tcpdump} is used to collect ICMP replies with timestamps. The average delay between packets is used as the only feature to classify the applications into CPU intensive, I/O intensive, or passive applications. A fuzzy neural classifier (NFC) provides the distance metric for clustering ground truth data into clusters and then using membership functions to separate future traffic patterns. Results show that high CPU load leads to responsive nodes, and I/O processes get bottle-necked by memory constraints leading to delayed network traffic. The study provides a case for future work in distinguishing between foreground and background traffic to identify operating system and memory characteristics by employing more features in classification.

Alan et al. \cite{24} consider application launch time network traffic, the first $n$ packets, as an application identifier, as they contain information with minimal user interference that is characteristic of the app. The application traffic is captured using  \texttt{tcpdump} at AP when a USB-Ethernet adapter connects the smartphone to an AP. 1595 apps on 4 Android devices and network traffic of 86,109 app launches were captured. ADB commands were used to install and uninstall the apps from the device, separately for training and testing datasets. This process was repeated for each of the apps all phones. Multinomial and gaussian na\ "{i}ve Bayes classifiers trained on packet lengths outperform Jaccard similarity on burst length. The classifiers were trained using the first six sessions, and the 7th session was used for testing. The practical use of this methodology has several pain points. The ability of such a framework to distinguish packet launch information from standard traffic on a monitored network is decisive for its usability in the absence of side-channel data. The dataset needs to be updated regularly, and models need to be retrained with samples from the latest versions of the apps to keep them accurate over time. Model accuracy is also affected by the operating system and build configuration statistics.

Shen et al. \cite{bigrams} use second-order Markov chains that output symbols corresponding to applications with feature bigrams as input. Application data and certificate packet lengths are used as features for the bigram, which the model uses to learn transition probabilities between states corresponding to application classes. Since Markov processes, like most methods, are prone to error for input data traces that do not often occur in the training dataset, the training cases are clustered to limit the range of values that the model has to learn in between the transitions. The model is shown to outperform similar Markovian process models. 

Lotfollahi et al. \cite{dpkt} and their previously discussed framework was also applied to the application identification paradigm of the ISCX dataset and achieved F-measure, precision, and accuracy of 0.98. Performance for both application and class of service problems fared better in competition with stacked autoencoders.

\subsection{Detection of activity inside an application}

\begin{table*}[h]
\centering
\caption{Activity Identification}
\label{activityTable}
\begin{tabular}{|c|c|c|p{2.8cm}|p{1.3cm}|p{2cm}|p{2cm}|p{2cm}|}
\hline
\multirow{2}{*}{S.No} & \multirow{2}{*}{Ref.} & \multicolumn{3}{c|}{Data Gathering Technique} & \multicolumn{3}{c|}{Methodology} \\ \cline{3-8}
& & Generation & Capturing & Tagging & Input & Algorithm & Feature \\ \hline
1. & Coull \& Dyer \cite{4} & Automated & At VPN & Automated & TCP packets & Lookup table & Packet length \vspace{5mm} \\ \hline
2. & Conti et al. \cite{41} & Manual & At server in a monitored network & Manual & TCP flow & Random forests, agglomerative hierarchical clustering & Packet length, cluster size \\ \hline
3. & Fu et al. \cite{17} & Manual & At virtual AP & Manual & TCP flow & Hidden Markov models, $k$-means clustering & Packet length, time delay \\ \hline
4. & Park \& Kim \cite{10} & Manual  & Between mobile and AP & Manual & Packet sequence & Random forests, agglomerative hierarchical clustering & Cluster statistics \\ \hline

\end{tabular}
\end{table*}
Limited methods are found to identify the actions performed by the applications or to identify specific feature sets used on the applications for promoting inferences from the network traffic. The notion of applying techniques from the previous two sections does not seem so far fetched, like predictive analysis of the class of services offered to the targeted user can help infer specific activities if application signatures or characteristics have already been identified. However, it is also natural to take the position that fine-grained actions will have equally finely isolated features in traffic at scales where labeling data meaningfully can be difficult, and expecting learning algorithms to have either good recall or precision is a substantial ask. 

Coull \& Dyer \cite{4} show that how an eavesdropper can get the information i.e., user actions, language, and length of the messages from instant messaging services i.e., WhatsApp, Apple iMessage by just observing the sizes of encrypted packets. Scripted actions for iMessage on iOS and OSX devices are used to generate labeled datasets of traffic on a monitored VPN. User action identification was among the intrusive objectives discussed, where learning algorithms were trained for each by 10-fold cross-validation. However, a hash map lookup based on packet length is showcased as most capable of identifying actions. This framework can be used very effectively for further information leakage in conjunction with an application identifier for iMessage.

Conti et al. \cite{41} investigates that at what extend, an eavesdropper can identify the actions performed on a mobile application by mobile users, by analyzing the encrypted traffic. Application traffic is routed to a back end cluster.  The framework limits itself to a selected sample of apps, for which traffic is filtered against a list of exempted IP addresses using the WHOIS protocol, and all other information is discarded. Flows are extracted from packets occurring in packet intervals defined by a timeout parameter. A supervised learning approach using random forests is pursued, and hence, a labeled dataset is required. Agglomerative hierarchical clustering is performed using dynamic time warping as a distance metric for sequences that allow matching elements while maintaining temporal order. Using scripted and timestamped actions on Facebook, Twitter, and Gmail apps, a dataset mapping flows to actions is thus curated with 220 sequences of fifty actions each for every application. For training a random forest classifier, only clusters that maximize the average F-measure are considered. Average precision, recall, and F-measure were above 0.97 for both Twitter and Facebook actions, and around the 0.85 marks for Gmail actions, suggesting that SSL/TSL encryption does not statistically encapsulate semantics of network traffic as efficiently as the syntax.

Fu et al. \cite{17} detail methodologies probed for identification of usage of individual services on mobile message applications. In-app usage can be used to profile user behavior, where data is collected by establishing a virtual access point to host the targeted smart-phone. Wireshark is used to intercept packets filtered by application layer protocols and to log statistics, where experiment labels are manually provided on the device. Traffic is segmented into timestamped sessions, that are further divided as individual dialogs. Packet-level statistics such as length and time delays in packets are used as features to create a set of vectorized dialogs. This feature set is used to predict the usage type prediction. Ensemble classifiers are developed, among which combining Hidden Markov Models and $k$-means clustering with either random forests or gradient boosted trees provide robust performance. 

Park \& Kim \cite{10} proposes a framework to infer performed tasks on KaKaoTalk, a mobile IM application. Actions like sending messages, adding friends, joining a chat group, and the like are considered. To analyze and capture the network traffic of the KakaoTalk application, the Android device is connected to a node mimicking a NAT Router and a packet capture application using the \texttt{libpcap} library. To collect data for training, 11 common activities on the messenger app were scripted, and 100 instances per activity are recorded. Flows are grouped using agglomerative hierarchical clustering using a DTW metric and labeled. 10-fold cross-validation was performed during training a random forest classifier, during which the number of clusters was optimized as a varying hyper-parameter. Using a random forest classifier with hierarchical clustering, all actions except for hiding a friend from the user's profile were classified accurately. However, the framework is notably agnostic to traffic statistics where all users are assumed to occur with equal probability, and packet features were not considered. Only text messages were considered, where multimedia messages may exhibit a very different network footprint.

\subsection{User and device fingerprinting}

Vanrykel et al. \cite{18} use a non-learning approach to user fingerprinting is described. A PC connected to a mobile device installs and performs scripted actions on applications and signals VPN servers to start and stop data capture. The data collected is parsed and filtered to extract HTTPS headers, packet timestamps, IP addresses, port numbers \emph{etc.} For example, various suitable identifiers from header bodies and URLs are then extracted to cluster the data stream belonging to each user. TCP timestamps and unique identifiers sent in HTTP traffic alone are shown to accurately cluster 57\% of users'users' mobile app sessions. 

St\"{o}ber et al. \cite{16} aim to identify configuration parameters of targeted smartphones based only on background traffic. The work assumes a passive agent capable of capturing encrypted wireless UMTS 3G data, intending to use network footprints of a target to identify their smartphone build. The attacker can access physical layer information and measure side-channel features in the 3G Network. Using \texttt{tcpdump}, 3G traffic is collected from 20 user devices for 8 hours with different combinations of 14 top apps installed with a restricted set of user interactions. Fingerprinting the specific device is achieved by distinguishing traffic bursts based on features like timing, size, and distribution. RMI (relative mutual information) of the features is calculated to measure the effectiveness of a feature in the problem. Packet size statistics were the most important with timing statistics holding a much smaller share of RMI, but all features were more informative than a random feature. The classifier used is a multi-class classifier using multiple binary $k$NN-SVM classifiers for test traffic data. It decides whether the traffic belongs to the device or not. Twenty-three features are noted for each burst for multiple apps on every smartphone, and the classifier is trained with 70\% data from one phone as positive data points and the rest from other devices to provide negative samples. The work claims that fingerprinting of all devices for 6 hours and using the classifier on monitored traffic for 15 minutes can achieve a classification accuracy of 90\%. 

Verde et al. \cite{23} explores methods to fingerprint the device and identify a particular user behind a NAT router based on his actions. The data used here is through Cisco's NetFlow protocol. The experiment is done on a metropolitan Wi-Fi network with a NetFlow enabled probe with the ISP. Five distinct users out of a possible two hundred thousand subscribers to the Network provided an identification test. Feature vectors are generated out of NetFlow records, which are used to train multiple hidden Markov models in parallel to model and filter inputs before performing a final classification. The HMM using which the model offers maximal F-measure is chosen. The average area under the ROC curve for the model across all users was optimal at 0.96 when using random forests as the final classification module. 

\begin{table*}[h]
\begin{center}
\caption{User and Device Fingerprinting}
\label{fprintTable}
\begin{tabular}{|c|c|c|p{2.5cm}|c|p{2cm}|p{2.2cm}|p{2.2cm}|}
\hline
\multirow{2}{*}{S.No} & \multirow{2}{*}{Ref.} & \multicolumn{3}{c|}{Data Gathering Technique} & \multicolumn{3}{c|}{Methodology} \\ \cline{3-8}
&  & Generation & Capturing & Tagging & Input & Algorithm & Features \\ \hline
1. & Vanrykel et al. \cite{18} & Scripted & At VPN & Automated & Application layer payload & Identifier matching (non-learning) & HTTPS identifiers \\ \hline
2. & St\"{o}ber et al. \cite{16} & Manual & At device & Manual & TCP flow& Ensemble classifier ($k$NN, SVM) & Packet length and timing statistics \\ \hline
3. & Verde et al. \cite{23} & Manual & At ISP using NetFlow protocol & Manual & Ordered bidirectional flow & Hidden Markov models \& random forests & Packet sequences\\ \hline
\end{tabular}
\end{center}
\end{table*}

\subsection{Operating system fingerprinting}
This is similar to the previous problem but presents different points of contention: network footprints of the same OS may vary strongly on different build configurations, and applications built on cross-platform libraries will colour the footprint in a similar manner across classes. For particular platforms like iOS or Chrome OS, effects caused by different configurations could be manageable because of the stronger coupling of hardware and software in the market. \\
Malik et al. \cite{5} describe an approach for which data is captured at a wireless access point using Wireshark. A standard ping utility or passive analysis of TCP/IP packets exchanged between the servers techniques is used. Ping flood testing is used to derive a probability distribution for the inter-packet spacing (IPS) of the ICMP replies captured from pinging the mobile devices. Similar distributions are also calculated for an active experiment with the mentioned devices connected to a monitored access point. Random forests are trained by 10-fold cross validation to identify the OS using statistical features of the mentioned distributions. Confusion matrix results suggest that the standard deviation and variance of ICMP replies are important features in classifying the dataset. Packet loss is also a significant factor that is characteristic of the OS and strongly affects the features monitored by the model.

Aksoy et al. \cite{aksoy} undertake OS identification of a target smartphone from frequency spectrum analysis of encrypted traffic on a wireless network. A dataset was generated using a controlled experimental network, that was used to collect ICMP replies from the most visited websites at the time using \texttt{traceroute}, and traffic generated during streaming and video chat on multiple machines for separate operating systems. A third of the data is isolated for testing, while the rest is used to train a classifier on feature sets selected by a genetic algorithm optimisted for feature extraction. Classifiers used for the experiment predominantly showed increased accuracy when boosted by the GA. 

Chen et al. \cite{21}, traffic is captured in the form of packet traces from Android, iOS and Windows Phone devices along with traces from a CRAWDAD dataset. Probabilistic learning using a na\"{i}ve Bayes classifier is compared with decision trees and linear regression. For the Bayesian classifier, the probability of the device using a specific OS as well as probability of tethering for different feature vectors is calculated. Selected features are used and evaluated in the experiment. The probabilistic classifier performs better compared to the other two classifiers by F-measure and recall, where all classifiers exhibit similar precision. 

Coull \& Dyer \cite{4} also address discerning the operating system (iOS or OSX) for iMessage traffic, where na\"{i}ve Bayes classifiers showcase perfect classification accuracy. The optimal number of required packets was probed, and using just five packets peak accuracy can be achieved. 
\begin{table*}[h]
\centering
\caption{Operating System Identification}
\label{osTable}
\begin{tabular}{|c|c|p{2.2cm}|p{2.2cm}|c|p{1cm}|p{2cm}|p{2.5cm}|}
\hline
\multirow{2}{*}{S.No} & \multirow{2}{*}{Ref.} & \multicolumn{3}{c|}{Data Gathering Technique} & \multicolumn{3}{c|}{Methodology} \\ \cline{3-8}
 &  & Generation & Capturing & Tagging & Input &Algorithm & Features\\ \hline
1. & Malik et al. \cite{5} & Scripted & At mobile \& AP & Automated & Ping flood response & Random forests & Average ICMP reply IPS \\ \hline
2. & Aksoy et al. \cite{aksoy} & Manual & b/w Mobile and AP & Manual& ICMP replies & Random forests, J48, OneR  & Selected by a genetic algorithm \\ \hline
3. & Chen et al. \cite{21} & Manual lab trace + CRAWDAD datasets & At access point (for lab trace) & Manual & TCP/IP packets & Na\"{i}ve Bayes classifier & TCP/IP statistics, machine clock frequency \\ \hline
4. & Coull \& Dyer \cite{4} & Automated & At VPN & Automated & TCP packets & Na\"{i}ve Bayes classifier & Packet length \\ \hline
\end{tabular}
\end{table*}
\subsection{Other prediction objectives}
Work by Coull \& Dyer \cite{4} has been mentioned before for discerning operating systems and user actions from iMessage traffic, and also explores message attribute extraction. It also shows that message language can be effectively predicted using a multinomial na\"{i}ve Bayes classifier, and linear regression can be used to approximate message length using payload length as a feature. Rich metadata about a user and their social network can hence be absorbed from packet statistics with high accuracy. 

Zhou et al \cite{2} use shared information in Android 4 applications to make privacy breaches. A malicious application with no permissions is used to test information leakage in resources shared between apps that monitors network traffic. The traffic signatures of all the apps are first learnt by analysis using Shark. Further, the application reads texttt{tcp\_rcv} and \texttt{tcp\_snd} values and compares the captured traffic with user activity signatures. Once the application being used is known the identity, location and even driving route preferences of the target can be known by further analysis.

\section{Countermeasures against NTA} \label{counter}
Statistical analysis of network communications seems to be enough to leak information. However, certain techniques can protect data privacy against highly efficient learning models to perform NTA on encrypted data.

A simple mechanism to thwart the encrypted traffic analysis is to make the size of the packets the same by padding them with some extra information and send the packets at a fixed interval of time. This helps reduce the information leakage about the traffic but incurs overhead and, consequently, affects the network protocols' performance and efficiency. However, such a padding scheme creates an imbalance between the privacy of the user and the performance of the network protocols. Indeed, all encrypted packets can be padded with the equal size of the maximum transmission unit (MTU), but for some networking protocols, such padding is not considered a suitable solution. \cite{liberatore2006inferring} shows that the per-packet padding mechanism helps reduce accuracy to 8\% while increasing the traffic volume to achieve the identification of encrypted HTTP streams. \cite{wright2008spot} used a padding mechanism to make the packet length multiple of 128, 256, and 512 bits block, which results in increasing overhead in a sequence but reducing recall and precision in a sequence. This padding mechanism is used in standardized security protocols, i.e., TLS, IPSec, SSh, \emph{etc.} Thus TLS, SSH, IPSec, and other security standards combine encryption and padding mechanisms. Padding makes it difficult for an attacker to attack based on the packet/message length analysis. However, it is found that hiding the packet length is not sufficient. TLS protocol is not found suitable for hiding the traffic patterns \cite{liberatore2006inferring}. There are some classifiers i.e., naive Bayes and VNG++ which don't use the packet length directly, instead, they use some or all following information; total bandwidth, burst size and overall time.

Traffic morphing technique \cite{28} is a defense mechanism found suitable to thwart the statistical-based encrypted traffic analysis by modifying the features of the packets in order to make the classes of traffic same by padding and truncating the packets. This technique changes the packet size and increases the number of packets by inserting some dummy traffic. Authors in \cite{28} use convex optimization techniques to perform real-time traffic modification with less overhead than padding.  This technique is applicable only for the traffic analysis techniques considered packet size as a feature. Traffic morphing technique is found suitable to create the balance between then privacy and network protocols' performance.

A study done by \cite{26} shows that these countermeasures against traffic analysis are found vulnerable to exploit the traffic features; hence, these countermeasures are not found effective. \cite{26} combines the discussed countermeasures and make a better scheme called as Buffered Fixed-Length Obfuscation (BuFLO). This scheme helps in reducing the success rate of information retrieval from attacks and provides better security but at a high bandwidth cost. Under this scheme, all packets, including dummy packets, should be of the same size and sent at a fixed interval. BuFLO helps in preventing the timing attacks as well by fixing the transmission rate of packets. \cite{cai2014systematic} proposed a better defense mechanism called Tamaraw, which is an extension of BuFLO. To design a better defense mechanism, it is essential to know which traffic feature is leaking the most information. Tamaraw uses different parameters to pad the packets than BuFLO with the consideration of packet direction. Authors in \cite{cai2014systematic} showed that Tamaraw provides better security than BuFLO with less wastage of bandwidth. These defense mechanisms can be applied against the traffic generated by mobile devices and non-mobile devices \cite{29}.

A defense mechanism called as Walkie-Talkie \cite{wang2017walkie} is presented against website fingerprinting with limited time and bandwidth overhead. Walkie-Talkie communicates in half-duplex mode instead of usual full-duplex mode by modifying the client's browser molded the traffic into a burst sequence. Similarly, to protect the attacker from knowing the password length in SSH (discussed in section 4) server can send some dummy packets if echo mode is off. However, this solution is not good to prevent timing attacks. A random amount of delay should be introduced between each keystroke to prevent timing attacks. Another way to prevent leakage of timing information is to send the traffic at a constant rate from both client and server-side \cite{song2001timing}.

\section{Conclusions} \label{conclusion}
   Current privacy provisions on the network heavily depend upon cryptography. However, it has been found that some information can still leak. Useful statistical features lie in the traffic flow that could be suitably explored with NTA. Nevertheless, merely the information related to the length and the inter-arrival time of the captured traffic, along with side-channel clues, is sufficient to extract several critical information about the traffic and profile the mobile user to whom that traffic belongs. 

   In this review, we focused on the research works involving the machine learning techniques for performing analysis of encrypted data generated by mobile devices. To represent the state-of-the-art related to the study topic, we proposed a classification framework to categorize the existing works with respect to their objectives of performing NTA, the method adopted by them to generate ground truth to build learning models, and the specific approach taken by them to achieve their objective. We also discussed several differences between the mobile and non-mobile Internet traffic that influence the techniques to perform NTA of these classes, respectively, and the difficulties and solutions in performing NTA on encrypted data.  

   Finally, we discussed countermeasures that can be taken by the applications to thwart NTA. While state-of-the-art research looks promising on extracting information from encrypted mobile traffic, the challenge is to keep their accuracy intact even when the statistical noises have been introduced in the traffic to thwart the NTA. On the other hand, an application applying some countermeasures against NTA needs to satisfy requirements, such as computational efficiency, preservation of semantics, and bandwidth utilization. There is a trade-off between network efficiency and information hiding when customizing packet length, which has been a critical feature in the methods described. A similar trade-off can also be imagined between the complexity of semantic noising or packet timing customization, and the corresponding drop in network utilization, as packets will have to be queued and sent in a manner such that the receiver can process them with logical coherence and can handle or minimize packet loss.

\section{Acknowledgement}
This research was supported by the Center for Artificial Intelligence and Robotics (CAIR) lab of Defence Research and Development Organisation (DRDO), India, Bangalore under the CARS scheme.

\bibliography{mybibfile}

\end{document}